\documentclass[journal,twoside,web]{ieeecolor}
\usepackage{graphicx,amstext,amsmath,amssymb,color,psfrag,mathabx,cite,booktabs}
\usepackage[latin1]{inputenc}

\let\labelindent\relax \usepackage{enumitem} 
\usepackage{comment}
\usepackage{url}            

\newtheorem{prop}{Property}

\definecolor{blue}{rgb}{0.0, 0.0, 1.0}
\definecolor{orange}{rgb}{1.0, 0.5, 0.0}

\newcommand{\change}[1]{\textcolor{black}{#1}}

\usepackage{generic}
\usepackage{amsfonts}
\usepackage{algorithm,algorithmic}

\usepackage{textcomp}
\def\BibTeX{{\rm B\kern-.05em{\sc i\kern-.025em b}\kern-.08em
    T\kern-.1667em\lower.7ex\hbox{E}\kern-.125emX}}
\begin{document}

\newcommand{\Rl}[2]{\ensuremath{\mathbb{R}^{#1}_{#2}}}   
\newcommand{\R}{\ensuremath{\mathbb{R}}}
\newcommand{\Rlp}{\ensuremath{\mathbb{R}_{>0}}}
\newcommand{\Rlpc}{\ensuremath{\overline{\mathbb{R}}_{>0}}}
\newcommand{\Rlo}{\ensuremath{\mathbb{R}_{\geq 0}}}
\newcommand{\Rln}{\ensuremath{\mathbb{R}_{< 0}}}
\newcommand{\Zo}{\ensuremath{\mathbb{Z}_{\geq 0}}}
\newcommand{\Zp}{\ensuremath{\mathbb{Z}_{> 0}}}
\newcommand{\Np}{\ensuremath{\mathbb{N}_{> 0}}}
\newcommand{\N}{\ensuremath{\mathbb{N}}}                 
\newcommand{\Z}{\ensuremath{\mathbb{Z}}}
\newcommand{\data}{\ensuremath{\mathrm{d}}}

\definecolor{bleucit}{rgb}{0.1,0.4,0.8}
\newcommand{\bleucit}{\textcolor{bleucit}}
\newcommand{\postoyanbleucit}{\textcolor{bleucit}}

\newcommand\cp[1]{{`\emph{#1}'}}
\newcommand\blue[1]{\emph{{\color{blue}#1}}}

\newcommand{\cmark}{\ding{51}}%
\newcommand{\xmark}{\ding{55}}%

\newcommand{\dst}{\displaystyle}
\newcommand{\Linf}[1]{\ensuremath{\mathcal{L}^{#1}}}

\newcommand{\eg}{{\it e.g.}}

\newcommand{\Nesic}{Ne{\v{s}}i{\'c} }

\definecolor{blue_cv}{rgb}{0.09,0.35,0.78}

\newcommand{\KL}{\ensuremath{\mathcal{KL}}}
\newcommand{\K}{\ensuremath{\mathcal{K}}}
\newcommand{\Kinf}{\ensuremath{\mathcal{K}_{\infty}}}
\newcommand{\KK}{\ensuremath{\mathcal{KK}}}
\newcommand{\KN}{\ensuremath{\mathcal{KN}}}
\newcommand{\KKL}{\ensuremath{\mathcal{KKL}}}
\newcommand{\KLL}{\ensuremath{\mathcal{KLL}}}
\newcommand{\D}{\ensuremath{\mathcal{D}}}
\newcommand{\PD}{\ensuremath{\mathcal{PD}}}

\newcommand{\Cs}{\ensuremath{C_{\text{steady}}}}
\newcommand{\Ct}{\ensuremath{C_{\text{transient}}}}
\newcommand{\Ds}{\ensuremath{D_{\text{steady}}}}
\newcommand{\Dt}{\ensuremath{D_{\text{transient}}}}
\newcommand{\UtGpAS}{U$_{\text{t}}$GpAS}
\newcommand{\UtGAS}{U$_{\text{t}}$GAS}
\newcommand{\UjGpAS}{U$_{\text{j}}$GpAS}
\newcommand{\UjGAS}{U$_{\text{j}}$GAS}

\newcommand{\argmin}{\ensuremath{\text{argmin}\,}}
\newcommand{\interior}{\ensuremath{\text{int}\,}}
\newcommand{\dom}{\ensuremath{\text{dom}\,}}
\newcommand{\Span}{\ensuremath{\text{Span}}}
\newcommand{\avg}{\ensuremath{\text{avg}}}
\newcommand{\co}{\ensuremath{\text{co}\,}}
\newcommand{\coc}{\ensuremath{\overline{\text{dom}}\,}}
\newcommand{\ext}{\ensuremath{\text{ext}}}
\newcommand{\rge}{\ensuremath{\text{rge}\,}}
\newcommand{\esup}{\ensuremath{\text{ess.sup}\,}}

\newcommand{\sign}[1]{\ensuremath{\text{sign}{(#1)}}}
\newcommand{\sat}{\ensuremath{\text{sat}}}

\newcommand{\sinc}{\ensuremath{\text{sinc}}}
\newcommand{\nom}{\ensuremath{\text{nom}}}

\newcommand{\Tmati}{\ensuremath{T_{MATI}\,\,}}
\newcommand{\Tmasp}{\ensuremath{\mathrm{T_{MASP}}\,}}
\newcommand{\lc}{\ensuremath{\llbracket}}
\newcommand{\rc}{\ensuremath{\rrbracket}}

\newcommand{\norm}[1]{\ensuremath{\left\|{#1}\right\|}}
\newcommand{\ip}[2]{\ensuremath{\left\langle #1, #2\right\rangle}}
\newcommand{\cb}[1]{\ensuremath{\overline{\mathbb{B}}_{\mbox{\scriptsize $#1$}}}}                              
\newcommand{\ob}[2]{\ensuremath{\mathbb{B}_{\mbox{\scriptsize $#1$}}\ensuremath{\left( #2\right)}}}   
\newcommand{\df}{\ensuremath{\stackrel{\mbox{\tiny $\mathrm{def}$}}{=}\:}}                                             
\newcommand{\myint}[4]{\ensuremath{\int_{#1}^{#2}#3\;\mathrm{d}#4}}
\newcommand{\Mm}{\ensuremath{\:\stackrel{\rightarrow}{\scriptstyle{\rightarrow}}\:}}
\newcommand{\bm}[1]{\ensuremath{\mathbf{#1}}}
\newcommand{\hs}[1]{\hspace*{#1 em}}
\newcommand{\qa}{\ensuremath{\mathcal{Q}_{A}}}%
\newcommand{\mc}[1]{\ensuremath{\mathcal{#1}}}
\newcommand{\di}{\ensuremath{\mathcal{D}_{i}}}

\newcommand{\HS}{\ensuremath{\mathcal{H}}}
\newcommand{\HSc}{\ensuremath{\mathcal{H}_c}}


%

\newcommand{\ie}{{\it i.e. }}

\newtheorem{exple}{Example}
\newtheorem{definition}{Definition}
\newtheorem{claim}{Claim}
\newtheorem{hypo}{Hypothesis}
\newtheorem{ass}{Assumption}
\newtheorem{proposition}{Proposition}
\newtheorem{fact}{Fact}
\newtheorem{lem}{Lemma}
\newtheorem{ex}{Example}
\newtheorem{thm}{Theorem}
\newtheorem{cor}{Corollary}
\newtheorem{pb}{Problem}
\newtheorem{rem}{Remark}
\newtheorem{sass}{Standing Assumption}


%
%
\newenvironment{rems}{\textit{Remarks. }}{\mbox{}\\[1ex]}
\newenvironment{eqn}[1][]{%
    \ifx&#1&\else\label{#1}\fi%
    \equation\begin{array}{rlllll}%
}{%
    \end{array}\endequation%
}

\title{A hybrid systems framework for data-based adaptive control of linear time-varying systems}
\author{Andrea Iannelli, \IEEEmembership{Member, IEEE}, Romain Postoyan, \IEEEmembership{Senior Member, IEEE}
\thanks{
This work was supported in part by the German Research Foundation (DFG) under Germany's Excellence Strategy - EXC 2075 - 390740016, and in part by the ANR under grant OLYMPIA ANR-23-CE48-0006.
}
\thanks{Andrea Iannelli is with the Institute for Systems Theory and Automatic Control, University of Stuttgart, Germany  (e-mail: andrea.iannelli@ist.uni-stuttgart.de).}
\thanks{Romain Postoyan is with the Universit\'e de Lorraine, CNRS, CRAN, F-54000 Nancy, France (e-mail: romain.postoyan@univ-lorraine.fr).}
}

\maketitle

\begin{abstract}
We consider the data-driven stabilization of  discrete-time linear time-varying systems. 
The controller is defined as a linear state-feedback law whose gain 
is adapted to the plant changes through a data-based event-triggering rule. To do so, we monitor the evolution of a data-based Lyapunov function along the solution. When this Lyapunov function does not satisfy a designed desirable condition, an episode is triggered to update the controller gain and the corresponding Lyapunov function using the last collected data. The resulting closed-loop dynamics hence exhibits both physical jumps, due to the system dynamics, and episodic jumps, which naturally leads to a hybrid discrete-time system. 
We leverage the inherent robustness of the controller and provide general conditions under which various stability notions can be established for the system. 
Two notable cases where these conditions are satisfied are treated, and numerical results illustrating the relevance of the approach are discussed.

\end{abstract}


\section{Introduction}
\label{sec:introduction}
\IEEEPARstart{T}{he} problem of designing controllers for dynamical systems only using data trajectories is an active area of research. Motivations include the increasing complexity of modern cyberphysical applications, the large availability of data and the time and cost savings potentially achieved by avoiding the modeling or system identification phase. Additionally, for systems changing over time, 
updating models during operation might not always be possible because by the time sufficiently many data for an accurate identification have been collected, the system dynamics could have substantially changed. In this work, we develop a framework for the synthesis of data-based adaptive controllers for  linear time-varying (LTV) systems with unknown state and input matrices, undergoing unpredicted time-variations. This problem poses two conceptual challenges. From a learning perspective, one should answer the questions of \emph{how} and \emph{when} to update the controller on the basis of data measured on-line.
From a technical perspective, proving closed-loop properties of the nonlinear time-varying interconnection between the plant and such an adaptive controller requires careful modeling and analysis steps relying on general, interpretable conditions. To address these challenges, we propose to cast this problem as an instance of hybrid (discrete-time) dynamical systems \cite[Section 4]{sanfelice-teel-aut10} and blend concepts from the event-triggered control literature e.g., \cite{Heemels-Johansson-Tabuada-cdc12} with recent results on data-based control \cite{de-persis-tesi-tac2020}. 

In the context of unknown LTI systems, the idea of designing a static state-feedback controller such that the closed-loop dynamics enjoys a favourable Lyapunov property can be traced back to \cite{de-persis-tesi-tac2020,van-waarde-et-al-tac2020(data)}. Here, stabilizing controllers are obtained by solving linear matrix inequalities (LMI) formulated in terms of suitable data matrices containing state and input trajectories collected offline, that is prior to the controller deployment. 
Whereas most of the approaches proposed in the literature use offline data and thus the resulting controller is fixed during operation, a few works studied \emph{adaptive} data-driven controllers, whereby the control law is modified during operation. In \cite{rotulo-et-al-aut22} the authors extended \cite{de-persis-tesi-tac2020} to the class of switched linear systems by using on-line collected data. The control update is performed at every time instant and the  stability guarantees rely on dwell-time conditions. 
The same class of systems was also studied in \cite{Eising_CDC22_DDswitched}, where it is assumed that data can be collected for each mode in an initialization phase, and the on-line task is to detect the active mode and then deploy the associated precomputed stabilizing feedback gain.
In \cite{Liu_CDC2023} 
an on-line approach is proposed for the more general class of LTV systems, 
and practical stability guarantees are obtained by implementing a periodic update 
based on the assumed rate of change of the system matrices. 
Data-driven control of LTV systems is also studied in \cite{nortmann-mylvaganam-tac23}, where 
finite horizon guarantees are derived by assuming that the data collected offline describe the behavior of the system during operation. 
To the best of the authors' knowledge, there is no prior work that considers stabilizing  data-driven adaptive controllers for general LTV systems where on-line data are used not only to update controllers, but also to decide on when the update should be performed. 

We propose to model the problem as a hybrid discrete-time system in the sense of \cite[Section 4]{sanfelice-teel-aut10}, which exhibits two types of jumps that are either due to the system dynamics or to the triggering of a new episode. By episode, we mean the update of the controller, chosen here as a linear state-feedback gain, based on the last collected state and input data. This modeling choice is justified by two facts. First, working with a hybrid discrete-time model allows us to derive an autonomous description of the system at hand capturing all the involved variables, which is essential to proceed with a rigorous analysis. These variables include the plant state and all the control design variables like the controller gain, the data collected to design the latter, as well as the variables related to the triggering mechanism. Second, as mentioned above, these variables evolve over two different times: the physical time for the plant state and the number of episodes for the control design related variables. Thus, it is natural to parameterize the solutions according to these two (hybrid) times. 
We present novel results for this class of hybrid dynamical systems, whose applicability goes beyond the scope of this work (Section \ref{sect:background}).  We then introduce all the variables needed to describe the overall closed-loop model and thus to obtain an autonomous hybrid discrete-time model 
(Section \ref{sect:problem-statement}). The objective is then to  design both the controller gain update rule as well as the episode-triggering law for the closed-loop system to guarantee stability of the origin under general conditions. 
We provide an explicit construction of the controller update rule building on the work in \cite{de-persis-tesi-tac2020} and extending it to our new problem setting (Section \ref{sect:controller-design}). 
Afterwards, we design the episode-triggering rule by exploiting a data-driven Lyapunov function obtained in the controller gain update step (Section \ref{sect:episode-triggering}). 
The trigger rule consists of monitoring whether this Lyapunov function satisfies a desired decay rate along the solution, similarly to \cite{Eising_CDC22_DDswitched}. When this is not the case, it means that the time-variations undergone by the plant require a new controller to be designed and thus a new episode to be triggered. Similar triggering rules have been proposed in e.g., \cite{Mazo-Anta-Tabuada-Aut10,Wang-Lemmon-aut11,maass-et-al-tac2023(etc),de-persis-et-al-tac2023(etc-data)}, where the set-up is different as a triggering instant refers there to a sampling instant, and not to a controller and Lyapunov function update, which leads to major modeling, analysis and design differences. 
In Section \ref{sect:stability-guarantees} we exploit the previous constructive results to establish sufficient conditions under which stability properties hold. 
A feature of these conditions is their generality, which allows covering a range of scenarios without explicitly imposing restrictions on the (variations of the) plant matrices. The price to pay is that these conditions may be difficult to check a-priori without additional knowledge of the plant. 
We therefore derive more specific conditions that relate instances of the variations of the system matrices to the stability guarantees ({as the ones verified in one of the numerical examples). 
Section \ref{sect:numerical-illustration} finally provides a numerical illustration of the proposed approach and compares it with the time-triggered controller from \cite{Liu_CDC2023}. 


We finally observe that, compared to related works mentioned above \cite{Liu_CDC2023,rotulo-et-al-aut22,Eising_CDC22_DDswitched}, the presented stability guarantees rely on general conditions that do not explicitly require knowledge on the system matrices, and the controller gain updates only occur when needed and not in a prefixed manner. 
Moreover, although recent works have used event-triggered concepts in the context of data-driven control e.g., \cite{qi2022data,wang-et-al-tc23,de-persis-et-al-tac2023(etc-data)}, the set-up here is very different. In these references, the control law is fixed once for all and the triggering rule is used to define the time at which sensor or actuator information is updated. 
Here, communication between the plant and the controller occurs at every physical time step and the triggering instant is the mechanism by which the controller gain is adapted. 
These key differences require completely different methodological tools.  
 

\section{Notation}\label{sect:notation} 

The symbol $\R$ denotes  the set of real numbers, $\Rlo:=[0,\infty)$, $\Rlp:=(0,\infty)$, $\Z$  the set of integers, $\Z_{\geq p}:=\{p,p+1,\ldots\}$ with $p\in\Z$ and $\emptyset$ the empty set. Given $k_1, k_2 \in \Zo$ with $k_1<k_2$, we denote by $[k_1,k_2]$ the finite set $\{k_1, k_1+1,.., k_2-1, k_2\}$. Given $n\in\Z_{\geq 1}$, we denote by $\mathbb{S}^n$ the set of real symmetric matrices of dimension $n$ and by $\mathbb{S}^n_{\succeq 0}$ and $\mathbb{S}^{n}_{\succ 0}$ the set of real symmetric positive semi-definite and positive definite matrices of dimension $n$, respectively. When $S\in\mathbb{S}_{\succ 0}^{n}$ with $n\in\Z_{\geq 1}$, its smallest and largest eigenvalues are denoted $\lambda_{\min}(S)$ and $\lambda_{\max}(S)$, respectively. The determinant of a real square matrix is denoted $\text{det}(\cdot)$, and its spectral norm by $\|\cdot\|$. 
We  use $\left[\begin{smallmatrix}
A & B \\
\star & C
\end{smallmatrix}\right]$ in place of $\left[\begin{smallmatrix}
A & B \\
B^\top & C
\end{smallmatrix}\right]$ for the sake of convenience. Given square matrices $A_1,\ldots,A_n\in\R^{n\times n}$ with $n\in\Z_{\geq 1}$, we use $\text{diag}(A_1,\ldots,A_n)$ to denote the block diagonal matrix  of dimension $n^2\times n^2$ whose block diagonal components are $A_1,\ldots,A_n$. 
The symbol $\otimes$ stands for the Kronecker product for matrices. 
Given $n, T\in \Z_{\geq 1}$,  $\mathcal{N}_{n,T}$ is the map from $\R^{n\times T}$ to $\R^{nT\times T}$ such that, for any $Z=[z_1,\ldots,z_T] \in \R^{n\times T}$, 
$
\mathcal{N}_{n,T}(Z)=\left[\begin{smallmatrix}
z_0 & 0& \ldots & 0 \\    
0 & z_1& \ldots & \vdots \\    
\vdots & \ddots & \ddots & \vdots \\    
0 & \cdots & \cdots & z_{T-1}  
\end{smallmatrix}\right]$, 
and we write $\mathcal{N}(Z)$ when $n$ and $T$ are clear from the context. Given $n,p,q\in\Z_{\geq 1}$ with $q\geq 2$ and a matrix $M=(M_{ij})_{(i,j)\in[1,n]\times[1,p]}\in\R^{n\times p}$, $M_{q:p}$ stands for the matrix $(M_{ij})_{(i,j)\in[1,n]\times[q,p]}\in\R^{n\times (p-q+1)}$ that is, the (thinner) matrix obtained by truncating
the first $q-1$ columns of matrix $M$. 
We use $I_N$ to denote the identity matrix of size $N\in\Z_{\geq 1}$, and $\mathbf{1}_N $ for the vector of $\R^{N}$, whose elements are all equal to $1$. 
The notation $(x,y)$ stands for $[x^{\top},\,y^{\top}]^{\top}$, where  $(x,y)\in\R^n\times\R^m$ and $n,m\in\Z_{\geq 1}$. The symbols $\lor$ and $\land$ stand for the ``or'' and ``and'' logic operators, respectively. 
The Moore-Penrose pseudoinverse of a real matrix $M$ is denoted by $M^\dagger$.

When we write $G:\mathcal{S}_1 \rightrightarrows\mathcal{S}_2$ for some sets $\mathcal{S}_1$, $\mathcal{S}_2$, it means that $G$ is a set-valued map from $\mathcal{S}_1$ to $\mathcal{S}_2$. We consider $\K$, $\K_\infty$ and $\KL$ functions as defined in \cite[Section 3.5]{Goebel-Sanfelice-Teel-book}, and we say that a continuous function from $\Rlo\to\Rlo$ is of class $\mathcal{L}$ if it is non-increasing and converges to $0$ at infinity. 
We also write $\beta\in\exp-\KL$ when $\beta:\Rlo\times\Rlo\to\Rlo$ and there exist $c_1\geq 1$ and $c_2>0$ such that $\beta(s,t)=c_1 e^{-c_2 t}s$ for any $(s,t)\in\Rlo\times\Rlo$. Given a function $\alpha:\Rlo\times\mathcal{S}\to\Rlo$ with $\mathcal{S}\subset\R^n$ for some $n\in\Z_{\geq 1}$, that is invertible in its first argument, we denote the corresponding inverse (with respect to the first argument) as $\alpha^{-1}(\cdot,z)$ for any $z\in\mathcal{S}$. Finally, for a set $S$, a (set-valued) map $f:S\rightrightarrows S$ and $N\in\Z_{\geq 1}$, $f^{N}$ stands for the $N^{\text{th}}$ composition of $f$ with itself.  

\section{Background}\label{sect:background}

This section provides the background material on the adopted hybrid modeling formalism from \cite{sanfelice-teel-aut10}. After having recalled the considered notion of solutions and given new conditions to ensure their completeness, various stability definitions are stated together with novel sufficient conditions to verify them. 
These results are of independent interest as their scope goes beyond the data-driven control problem addressed in the sequel. 

\subsection{Hybrid discrete-time systems}

We consider dynamical systems given by  
\begin{equation}
\left\{\begin{array}{rlllll}
q^+  =  f(q) & & q\in \mathcal{C} \\
q^+  \in  G(q) & & q\in \mathcal{D},
\end{array}\right.\label{eq:sys-background}
\end{equation}
where $\mathcal{C}\subset \R^{n_q}$, $\mathcal{D}\subset \R^{n_q}$, $f:\mathcal{C}\to\R^{n_q}$ and $G:\mathcal{D}\rightrightarrows\R^{n_q}$ a set-valued map with $n_q\in\Z_{\geq 1}$; see \cite[Section 4]{sanfelice-teel-aut10}. Solutions to (\ref{eq:sys-background}) exhibit two types of jumps depending on whether they belong to set $\mathcal{C}$ or to set $\mathcal{D}$  as formalized below. In the context of this paper, $q$ is the concatenation of the plant state vector and some auxiliary variables,  $\mathcal{C}$ is the region of the state space where the system is operated, and $\mathcal{D}$ is the region of state space where a new controller is learned 
and, using the terminology adopted in the rest of the work, a new episode is triggered.  
We therefore parameterize solutions to (\ref{eq:sys-background}) using two different times as in \cite[Section 4]{sanfelice-teel-aut10} to  distinguish jumps that are due to the dynamics  on $\mathcal{C}$ and jumps that are  due to  the dynamics defined on $\mathcal{D}$. This is the reason why we call system (\ref{eq:sys-background}) a \emph{hybrid discrete-time system}. 

\subsection{Solution concept}\label{subsection:background-solution}

We adopt the notion of solution advocated in \cite[Section 4]{sanfelice-teel-aut10}.  A subset $E\subset\Zo\times\Zo$ is a \emph{compact discrete time domain} if $E=\bigcup_{j=0}^{J-1}\bigcup_{k=k_j}^{k_{j+1}}(k,j)$ for some finite sequence $0=k_0\leq k_1\leq \ldots\leq k_{J}$, $k_j\in\Zo$ for every $j\in[0,J]$ with $J\in\Z_{\geq 1}$, and it is a \emph{discrete time domain} if for any $(k,j)\in E$, $E\cap([0,k]\times[0,j])$ is a compact discrete time domain \cite[Definition 4.1]{sanfelice-teel-aut10}. A function $q:\dom q\to \R^{n_q}$ is a \emph{discrete arc} 
if $\dom q$ is a discrete time domain \cite[Definition 4.2]{sanfelice-teel-aut10}. We are ready to define  solutions to  (\ref{eq:sys-background}). 

\begin{definition}[$\!\!${\cite[Definition 4.3]{sanfelice-teel-aut10}}]\label{def:solution} A discrete arc $q:\dom q\to \R^{n_q}$ is a \emph{solution} to system (\ref{eq:sys-background}) if 
\begin{enumerate}
\item[(i)] for all $k,j\in\Zo$ such that $(k,j),(k+1,j)\in\dom q$, $q(k,j)\in \mathcal{C}$ and $q(k+1,j)=f(q(k,j))$,
\item[(ii)] for all $k,j\in\Zo$ such that $(k,j),(k,j+1)\in\dom q$, $q(k,j)\in \mathcal{D}$ and $q(k,j+1)\in G(q(k,j))$. \hfill $\Box$
\end{enumerate} 
\end{definition}

Definition \ref{def:solution} essentially means that a solution to (\ref{eq:sys-background}) jumps according to: (i) the map $f$ when it lies in $\mathcal{C}$; (ii) the set-valued map $G$ when it lies in  $\mathcal{D}$; (iii) either $f$ or $G$ when it lies  in $\mathcal{C}\cap \mathcal{D}$. 
We note that, as mentioned above, solutions to (\ref{eq:sys-background}) are parameterized by two times: $k$ and $j$, which in our setting respectively correspond to the physical time and the episodic time that is,  the number of episodes experienced so far by the solution. We call a pair $(k,j)$ a \emph{hybrid time}.

We give a few other definitions relevant to our work. 
\begin{definition}\label{def:properties-solution-domains} A solution $q$ to (\ref{eq:sys-background}) is:
\begin{itemize}
\item \emph{maximal}  if there does not exist another solution $q'$ to (\ref{eq:sys-background}) such that $\dom q$ is a proper subset of $\dom q'$ and $q(k,j)=q'(k,j)$ for any $(k,j)\in\dom q$;
\item \emph{complete} if $\dom q$ is unbounded;
\item $k$-\emph{complete} if $\sup\{k\in\Zo\,:\,\exists j\in\Zo,\,\,(k,j)\in\dom q\}=\infty$.\hfill $\Box$
\end{itemize}    
\end{definition}

We provide below  conditions for any maximal solution to (\ref{eq:sys-background}) to be $k$-complete.

\begin{proposition}\label{prop:completeness-background} Consider system (\ref{eq:sys-background}), the following holds.
\begin{enumerate}
\item[(i)] Any maximal solution  is complete if and only if $f(\mathcal{C})\cup G(\mathcal{D}) \subset \mathcal{C}\cup\mathcal{D}$. 
\item[(ii)] Any maximal solution is $k$-complete if  $f(\mathcal{C})\cup G(\mathcal{D}) \subset \mathcal{C}\cup\mathcal{D}$ and there exists $N\in\Z_{\geq 1}$ such that $G^{N}(\mathcal{D})\cap \mathcal{D}=\emptyset$. \hfill $\Box$
\end{enumerate}
\end{proposition}

\noindent\textbf{Proof:} (i) Suppose that any maximal solution to (\ref{eq:sys-background}) is complete. This implies that for any point $q$ in $\mathcal{C}$, $f(q)$ belongs to $\mathcal{C}\cup \mathcal{D}$,  and from any point $q$ in $\mathcal{D}$ and any $g\in G(q)$,  $g\in\mathcal{C}\cup \mathcal{D}$, otherwise we would have the existence of a solution, which would leave $\mathcal{C}\cup \mathcal{D}$ after one jump and would thus not be complete but this contradicts the made assumption. We have established that when any maximal solution to (\ref{eq:sys-background}) is complete, $f(\mathcal{C})\cup G(\mathcal{D})\subset \mathcal{C}\cup\mathcal{D}$.  On the other hand, when  $f(\mathcal{C})\cup G(\mathcal{D}) \subset \mathcal{C}\cup\mathcal{D}$, this implies the forward invariance of  $\mathcal{C}\cup \mathcal{D}$ for system (\ref{eq:sys-background}) in the sense that any solution to (\ref{eq:sys-background}) initialized in $\mathcal{C}\cup\mathcal{D}$ remains in this set for all future times. This property implies that any maximal solution to (\ref{eq:sys-background}) remains in $\mathcal{C}\cup \mathcal{D}$ and is thus complete: item (i) of Proposition \ref{prop:completeness-background} holds. 

(ii) 
Since $f(\mathcal{C})\cup G(\mathcal{D}) \subset \mathcal{C}\cup\mathcal{D}$, any maximal solution to (\ref{eq:sys-background}) is complete by item (i) of Proposition \ref{prop:completeness-background}. Suppose that there exists a maximal solution $q$ to (\ref{eq:sys-background}) that is not $k$-complete. Since $q$ is complete that means there exists $(k,j)\in\dom q$ such that $(k,j')\in\dom q$ for any $j'\geq j$. Consequently, for any $N\in\Z_{\geq 1}$, $G^N(\{q(k,j)\})\cap \mathcal{D}\neq \emptyset$. This implies, as $q(k,j)\in\mathcal{D}$, that  $G^N(\mathcal{D})\cap\mathcal{D}\neq \emptyset$ for any $N\in\Z_{\geq 1}$: this contradicts the fact that there exists $N\in\Z_{\geq 1}$ such that $G^N(\mathcal{D})\cap \mathcal{D}=\emptyset$. We have obtained the desired result by contradiction. \hfill $\blacksquare$


Lastly, we associate with any hybrid arc $q$ of (\ref{eq:sys-background}) and any hybrid times $(k,j)\in\dom q$, the sequence $k_i\in\Zo$, $i\in\{0,j\}$ such that: $0=k_0\leq k_1\leq \ldots \leq k_{j} \leq k$; and $\dom q\cap ([0,k]\times[0,j])=\big(\bigcup_{i\in[0,j-1]}[k_i,k_{i+1}]\times\{i\}\big)$ $\bigcup\big([k_j,k]\times\{j\}\big)$. In words, the $k_i$'s will be the physical times at which a new episode is triggered for solution $q$ in the setting of this work.



    

\subsection{Stability definitions}\label{subsect:stability-background}

We will investigate  stability of closed, unbounded sets of the form 
\begin{equation}\label{eq:set-A}
\mathcal{A}=\{q=(q_1,q_2)\in\mathcal{S}_{1}\times\mathcal{S}_{2}\,:\,q_1=0\},
\end{equation}
where $\mathcal{S}_{1}\subset\R^{n_{1}}$, $\mathcal{S}_{2}\subset\R^{n_{2}}$, $\mathcal{S}_1\times\mathcal{S}_2=\mathcal{C}\cup \mathcal{D}$, $n_{1},n_{2}\in\Z_{\geq 1}$ and $n_{1}+n_{2}=n_q$, for  system (\ref{eq:sys-background}).

\begin{definition}\label{def:stability} Consider system (\ref{eq:sys-background}) and set $\mathcal{A}$ in (\ref{eq:set-A}). We say that:
\begin{itemize}
\item $\mathcal{A}$ is \emph{stable} if for any $\varepsilon\in\Rlp$, for any $q_{2,0}\in\mathcal{S}_{2}$, there exists $\delta(\varepsilon,q_{2,0})\in\Rlp$ such that any solution $q$ with $|q_1(0,0)|\leq \delta(\varepsilon,q_{2,0})$ and $q_2(0,0)=q_{2,0}$ verifies $|q_1(k,j)|\leq \varepsilon$ for any $(k,j)\in\dom q$;
\item $\mathcal{A}$ is \emph{uniformly stable} if for any $\varepsilon\in\Rlp$, there exists $\delta(\varepsilon)\in\Rlp$ such that any solution $q$ with $|q_1(0,0)|\leq \delta(\varepsilon)$ verifies $|q_1(k,j)|\leq \varepsilon$ for any $(k,j)\in\dom q$;
\item $\mathcal{A}$ is \emph{globally attractive} if any complete solution $q$ verifies 
$|q_1(k,j)|\to 0$ as $k+j\to\infty$;
\item $\mathcal{A}$ is \emph{globally asymptotically stable (GAS)} if it is stable and globally attractive;
\item $\mathcal{A}$ is \emph{uniformly globally asymptotically stable (UGAS)} if there exists $\beta\in\KL$  such that for any solution $q$, $|q_1(k,j)|\leq \beta(|q_1(0,0)|,k+j)$ for any $(k,j)\in\dom q$;
\item $\mathcal{A}$ is \emph{uniformly globally exponentially stable (UGES)}  if there exists $\beta\in\exp-\KL$ such that for any solution $q$,   $|q_1(k,j)|\leq \beta(|q_1(0,0)|,k+j)$ for any $(k,j)\in\dom q$.
 \hfill $\Box$
\end{itemize}
\end{definition}

Definition \ref{def:stability} formalizes various stability notions and distinguishes whether the stability property is uniform or not. This distinction is justified in the context of this work  because we will be dealing with a set $\mathcal{A}$ of the form of (\ref{eq:set-A}), also called attractor,  that is closed but not bounded. For more insights on uniform vs non-uniform stability, see \cite[Examples 3.9-3.15]{Goebel-Sanfelice-Teel-book} in the general context of hybrid inclusions. 


We present next relaxed Lyapunov conditions to ensure the stability notions of Definition \ref{def:stability} for set $\mathcal{A}$ in (\ref{eq:set-A}) for system (\ref{eq:sys-background}). By relaxed conditions, we mean that the considered Lyapunov function candidate does not  need to strictly decrease along solutions to (\ref{eq:sys-background}) at every point of the state space outside of $\mathcal{A}$, consistently with e.g., \cite[Chapter 3.3]{Goebel-Sanfelice-Teel-book}.  This type of Lyapunov properties is very natural for the data-driven control problem we address as we will see. 

\subsection{Lyapunov conditions}\label{subsect:background-Lyapunov}

To prove the properties of Definition \ref{def:stability}, we will be constructing a Lyapunov function candidate $\mathcal{U}:\mathcal{C}\cup \mathcal{D}\cup f(\mathcal{C})\cup G(\mathcal{D})\to\Rlo$ verifying the next properties.
\begin{enumerate}
\item[(P1)] There exist  $\underline\alpha,\overline\alpha:\Rlo\times\mathcal{S}_2\to\Rlo$ of class-$\Kinf$ in their first argument such that $
\underline\alpha(|q_1|,q_2)  \leq   \mathcal{U}(q)  \leq   \overline\alpha(|q_1|,q_2)$ for any $q=(q_1,q_2)\in \mathcal{C}\cup \mathcal{D}$.
\item[(P2)] There exists $\nu_{c}:\mathcal{C}\to\Rlo$ such that
$\mathcal{U}(f(q)) \leq \nu_c(q) \mathcal{U}(q)$ for any $q\in \mathcal{C}$.
\item[(P3)] There exists $\nu_{d}:\mathcal{D}\to\Rlo$ such that 
$\mathcal{U}(g) \leq \nu_d(q) \mathcal{U}(q)$ for any $ q\in \mathcal{D}$ and any $g\in G(q)$.
\end{enumerate}

Given (P2)-(P3), we have the next property for function $\mathcal{U}$ along any solution to  (\ref{eq:sys-background}). 

\begin{lem}\label{lem:lyap-bound} Consider system (\ref{eq:sys-background}) and suppose (P2)-(P3) hold. Any solution $q$ to (\ref{eq:sys-background}) satisfies
\begin{eqn}\label{eq:lemma-lyap-bound}
\mathcal{U}(q(k,j)) \leq \dst \pi(q,k,j) \mathcal{U}(q(0,0)) & & \forall (k,j)\in\dom q,
\end{eqn}
with 
\begin{eqn}\label{eq:pi}
\overbrace{\prod_{j'=0}^{j-1}\prod_{k'=k_{j'}}^{k_{j'+1}-1}\nu_c(q(k',j'))\nu_d(q_2(k_{j'+1},j')) \prod_{k''=k_{j}}^{k-1}\nu_c(q(k'',j))}^\text{{\normalsize $\pi(q,k,j) := $}},
\end{eqn}
and $k_0,k_1,\ldots, k_{j}$ as defined at the end of Section \ref{subsection:background-solution}. 
\hfill $\Box$
\end{lem}
\noindent\textbf{Proof:} Let $q$ be solution to (\ref{eq:sys-background}), $(k,j)\in\dom q$. 
By (P2), we have 
$
\mathcal{U}(q(k_1,0))\leq \prod_{k'=0}^{k_1-1}\nu_c(q(k',0))\mathcal{U}(q(0,0))$. 
At $(k_1,1)$, 
$\mathcal{U}(q(k_1,1))\leq \nu_d(q(k_1,0)) \mathcal{U}(q(k_1,0))$,
therefore $\mathcal{U}(q(k_1,1))\leq \prod_{k'=0}^{k_1-1}\nu_c(q(k',0))\nu_d(q(k_1,0))$ $\times \mathcal{U}(q(0,0))$. We obtain the desired result by iterating the same argument until $(k,j)$.  
\hfill$\blacksquare$

Property (\ref{eq:lemma-lyap-bound}) is  not enough to conclude about stability properties for set $\mathcal{A}$ for system (\ref{eq:sys-background}), extra conditions are required that are formalized next.
\begin{thm}\label{theorem:stability} Consider system (\ref{eq:sys-background}) and suppose (P1)-(P3) are satisfied. The following holds.
\begin{enumerate}[label=(\roman*)]
\item If there exists $\vartheta:\Rlo\times\mathcal{S}_2\to\Rlo$ of class-$\Kinf$ in its first argument such that for any solution $q$, 
\begin{eqn}\label{eq:prop-stability-bound-pi-stability}
\overline\pi(q,k,j) \leq  \vartheta(|q_1(0,0)|,q_2(0,0)) &  \forall (k,j)\in\dom q,
\end{eqn}
with $\overline\pi(q,k,j):=\underline\alpha^{-1}\Big(\pi(q,k,j)\overline\alpha(|q_1(0,0)|,q_2(0,0)),$ $q_2(k,j)\Big)$, $\underline\alpha,\overline\alpha$ in (P1) and $\pi$ in (\ref{eq:pi}), then $\mathcal{A}$ is stable. In addition, 
\begin{enumerate}[leftmargin=1cm]
    \item[(i-a)] if $\vartheta(s,\cdot)$ is constant for any $s\in\Rlo$, then $\mathcal{A}$ is uniformly stable,
    \item[(i-b)] if for any complete solution $q$, $\overline\pi(q,k,j)\to 0$ as $k+j\to\infty$, then $\mathcal{A}$ is GAS.
\end{enumerate}
\item If there exists $\beta\in\KL$ (respectively, $\beta\in\exp-\KL$) such that  for any solution $q$, 
\begin{eqn}\label{eq:prop-stability-KL}
\overline\pi(q,k,j) \leq  \beta(|q_1(0,0)|,k+j) &  \forall (k,j)\in\dom q,
\end{eqn}
then $\mathcal{A}$ is UGAS (respectively, UGES).\hfill $\Box$
\end{enumerate}
\end{thm}
\noindent\textbf{Proof:} (i) Let $q$ be a solution to (\ref{eq:sys-background}).  By Lemma \ref{lem:lyap-bound}, $\mathcal{U}(q(k,j))\leq \pi(q,k,j) \mathcal{U}(q(0,0))$. Consequently, in view of (P1), $|q_1(k,j)|\leq\overline\pi(q,k,j)$ with  $\overline\pi$ defined in item (i), and thus $|q_1(k,j)|\leq\vartheta(|q_1(0,0)|,$ $q_2(0,0))$ by (\ref{eq:prop-stability-bound-pi-stability}). Set $\mathcal{A}$ is stable according to  Definition \ref{def:stability} as the required property holds by taking $\delta=\vartheta^{-1}(\varepsilon,q_{2,0})$ for any $\varepsilon>0$ and any $q_{2,0}\in\mathcal{S}_2$. Item (i-a) then follows. When, in addition to $\mathcal{A}$ being stable, $\overline\pi(q,k,j)\to 0$ for $k+j\to\infty$ when $q$ is complete, we have $|q_1(k,j)|\to 0$ as $k+j\to\infty$, as $|q_1(k,j)|\leq\overline\pi(q,k,j)$ for any $(k,j)\in\dom q$. This means $\mathcal{A}$ is globally attractive and thus GAS.  (ii) The desired result is obtained by exploiting similar bounds as above together with (\ref{eq:prop-stability-KL}). \hfill $\blacksquare$

We will exploit the conditions of Theorem \ref{theorem:stability} to establish properties for the data-driven control problem presented in the next section.

\section{Hybrid modeling and objectives}\label{sect:problem-statement} 

This section starts with the presentation of the plant model and the main goal of this work (Section \ref{subsect:plant}). Afterwards, all the variables needed in the control design are introduced (Section \ref{subsect:additional variables}), and the overall hybrid discrete-time model is then derived (Section \ref{subsect:hybrid-modeling}). We conclude this section with a formal statement of the design objectives (Section \ref{subsect:design-objectives}).

\subsection{Plant}\label{subsect:plant}
Consider the discrete-time LTV system
\begin{equation}
\begin{array}{rlll}
x(k+1) & = & A(k)x(k) + B(k)u(k),
\end{array}\label{eq:plant}
\end{equation}
where $x(k)\in\R^{n_x}$ is the state and $u(k)\in\R^{n_u}$ is the control input at time $k\in\Zo$, with $n_x,n_u\in\Z_{\geq 1}$. Time-varying matrices $A(k)$ and $B(k)$ are \emph{unknown} and take values in $\R^{n_x\times n_x}$ and $\R^{n_x\times n_u}$ for any $k\in\Z_{\geq 0}$, respectively.  

Our goal is to stabilize the origin of system (\ref{eq:plant}) despite the fact that $A(\cdot)$ and $B(\cdot)$ are unknown and time-varying. 
To address these two challenges in a direct data-driven fashion whereby only measured data are used, we propose using an event-triggered approach.  
Specifically, we apply the last constructed linear state-feedback law until it is outdated in the sense that a state-dependent condition related to some appropriately designed Lyapunov property is \emph{not} fulfilled. 
The violation of this condition triggers a new episode where, whenever this is possible, a new feedback law is computed based on the last collected data, and then applied to plant (\ref{eq:plant}). 
We then keep repeating these steps, and provide a closed-loop stability analysis of this nonlinear adaptive feedback interconnection by framing it as an instance of the hybrid system class described in Section \ref{sect:background}.

To model the problem as a hybrid discrete-time system, we need to introduce  auxiliary variables appearing in 
the design  of the feedback gains and the episode-triggering condition. 

\subsection{Auxiliary variables}\label{subsect:additional variables}

All the variables introduced in the sequel are summarized in Table \ref{tab:variables} together with their ``informal'' meaning and the set over which they are defined.

\begin{table*}[t]
\centering
\begin{tabular}{lll}
\toprule
$x$ & Plant state & $\R^{n_x}$ \\
$\kappa$ & Physical time counter & $\Zo$ \\
$\widehat X$ & Collection of the value of $x$ from $k-T$ to $k-1$ & $\R^{n_x\times T}$ \\
$X$ & Collection of the value of $x$ from $k-T+1$ to $k$ & $\R^{n_x\times T}$ \\
$U$ & Collection of the value of $u$ from $k-T$ to $k-1$ & $\R^{n_u\times T}$ \\
$K$ & Controller gain & $\R^{n_u\times n_x}$\\
$S$ & Lyapunov-like matrix & $\mathcal{S}\subset\mathbb{S}^{n_x}_{\succ 0}$\\
$F$ & Matrix involved in the  controller update rule & $\mathbb{S}^{n_x}_{\succ 0}$\\
$a_1,a_2$ & Scalar variables involved in the controller update rule & $[0,1]$, $\Rlo$\\
$\eta$ & Variable involved in the episode-triggering condition & $\mathcal{S}_\eta\subset\R^{n_\eta}$\\
$q$ & Concatenation of all the state variables, see (\ref{eq:q}) & $\mathcal{S}_q$ in (\ref{eq:q}) \\
$q_1$ & Plant state $x$ with the notation of Section \ref{subsect:stability-background} &  $\mathcal{S}_1=\R^{n_x}$ \\
$q_2$ & All the state variables but $x$, i.e., $(\kappa,\widehat X,X,U,K,S,F,a_1,a_2,\eta)$ &  $\mathcal{S}_2$ in (\ref{eq:set-A-H-notation-background}) \\
$\hat x$ & Value of state $x$ at the last physical time & $\R^{n_x}$\\
$\data$ & Concatenation of the time counter and data-matrices, i.e., $(\kappa,\widehat X,X,U)$ &  $\mathcal{S}_{\data}:=\Zo\times\R^{n_x\times T}\times\R^{n_x\times T}\times \R^{n_u\times T}$\\
\bottomrule 
\end{tabular}
\caption{Summary of the variables}\label{tab:variables}
\end{table*}

\subsubsection{Counter} 
We first introduce a counter\footnote{See Remark \ref{rem:counters} for a technical justification.} $\kappa\in\Zo$ of the physical time 
on which $A$ and $B$ depend in (\ref{eq:plant}). Hence, when no new episode is triggered, 
\begin{eqn}\label{eq:sys-timer-C}
\kappa^{+}  & = &  \kappa + 1,
\end{eqn}
and when a new episode is triggered, $\kappa$ keeps the same value, 
\begin{eqn}\label{eq:sys-timer-D}
\kappa^{+}  & = &  \kappa.
\end{eqn} 
As a result, by (\ref{eq:plant}), when no episode is triggered, 
\begin{eqn}\label{eq:plant-timer-C}
x^+ & = & A(\kappa)x + B(\kappa)u,
\end{eqn}
and, when a new episode is triggered,
\begin{eqn}\label{eq:plant-timer-D}
x^+ & = & x.
\end{eqn}

\subsubsection{Data variables}\label{subsubsection:data-variables}

To design a feedback controller with some desired properties for system (\ref{eq:plant-timer-C})-(\ref{eq:plant-timer-D}), we use at each episode-triggering time instant the values of the plant state $x$ and of the input $u$ over the last $T$ physical time steps, where $T\in\Z_{\geq 1}$ is a design parameter. We elaborate below on the choice of $T$  and on episodes that may be triggered before $T$ physical steps have passed in the sequel (Remarks \ref{rem:initial-data-variables} and \ref{rem_choice_T}).%

To model the data collection process, we introduce variables $\widehat{X},X\in\R^{n_x\times T}$ and $U\in\R^{n_u\times T}$, whose dynamics are, when no episode is triggered, 
\begin{eqn}\label{eq:sys-data-C}
\widehat{X}^{+}  & = &  [\widehat{X}_{2:T},x]\\
X^{+}  & = &  [X_{2:T}, A(\kappa)x+B(\kappa)u]\\
U^{+}  & = &  [U_{2:T}, u],
\end{eqn}
where we recall that $M_{2:T}$ is the matrix obtained by truncating
the first column of matrix $M$, and when a new episode is triggered,
\begin{eqn}\label{eq:sys-data-D}
(\widehat{X}^{+},X^{+},U^{+})  & = &  (\widehat{X},X,U).
\end{eqn}
Equations (\ref{eq:sys-data-C})-(\ref{eq:sys-data-D}) mean that, at time $(k,j)$ with $k\geq T$, $\widehat X$ and $U$ collect the values of 
$x$ and $u$ from physical time $k-T$ to $k-1$, respectively, and $X$ collects the value of $x$ from $k-T+1$ to $k$ as they remain unchanged when a new episode is triggered. 
The next lemma formalizes these claims considering system (\ref{eq:sys-timer-C})-(\ref{eq:sys-data-D}) as a hybrid system of the form of (\ref{eq:sys-background}) where set $\mathcal{C}$ corresponds to no episode triggering and $\mathcal{D}$ to episode triggering. 


\begin{lem}\label{lem:data-based-matrices} Given $T\in\Z_{\geq 1}$, for any solution $(x,\kappa,\widehat X,X,U)$  to (\ref{eq:sys-timer-C})-(\ref{eq:sys-data-D}) and any discrete arc\footnote{Although we did not define solutions to hybrid discrete-time systems with external inputs in Section \ref{sect:background}, in the following  $u$ will be a function of the state variables so that the adopted notion of solution in Definition \ref{def:solution} will apply.} $u$, for any $(k,j)\in\dom (x,\kappa,\widehat X,X,U,u)$  with  $k\geq T$, 
\begin{equation}
    \begin{array}{rllll}
    \widehat{X}(k,j) & = & \left[x(k-T,j_{k-T})\,\ldots\, x(k-1,j_{k-1})\right]\in\R^{n_x\times T}\\
    X(k,j) & = & \left[x(k-T+1,j_{k-T+1})\,\ldots\, x(k,j_{k})\right]\in\R^{n_x\times T}\\
    \mathcal{U}(k,j) & = & \left[u(k-T,j_{k-T})\,\ldots\, u(k-1,j_{k-1})\right]\in\R^{n_u\times T},
    \end{array}\label{eq:initial-data}
\end{equation}
where $j_{k'}\in\Zo$ is such that $(k',j_{k'})\in\dom (x,\kappa,\widehat X,X,U,u)$ for any $k'\in\Zo$. \hfill $\Box$    
\end{lem}

\noindent\textbf{Proof:} Let $(x,\kappa,\widehat X, X, U)$ be a solution to (\ref{eq:sys-timer-C})-(\ref{eq:sys-data-D}) with discrete arc  $u$. By (\ref{eq:sys-data-C}), 
$\widehat{X}(1,j_1) = [\widehat X_{2:T}(0,0),x(0,0)]$.
Then, 
\begin{equation}
\begin{array}{rclll}
\widehat{X}(2,j_2) & = & [\widehat X_{2:T}(1,j_1),x(1,j_1)]\\
& \vdots & \\
\widehat{X}(T,j_T) & = & [\widehat X_{2:T}(T-1,j_{T-1}),x(T-1,j_{T-1})]\\
& = & [x(0,0), x(1,j_j),\ldots,x(T-1,j_{T-1})].
\end{array}\label{eq-proof-lemma-data}
\end{equation}
By repeating these steps, we have for any $(k,j)$ in the solution domain  with $k\geq T$, $\widehat{X}(k,j)  =  [x(k-T,j_{k-T}), \ldots,x(k-1,j_{k-1})]$, 
which corresponds to the first line of (\ref{eq:initial-data}). We similarly obtain the other equations of (\ref{eq:initial-data}). \hfill $\blacksquare$

Lemma \ref{lem:data-based-matrices} implies that variables $\widehat X$, $X$ and $U$ are consistent with the data matrices commonly encountered for the data-driven control of LTI systems (see, e.g., \cite{de-persis-tesi-tac2020,van-waarde-et-al-tac2020(data)}) after $T$ physical time steps have elapsed. Contrary to the offline LTI system setting where these have been previously employed, it is essential here to include these data-based matrices in the state vector as they evolve with time and play a key role in the design of the adaptive controller.  In this way, the developed hybrid model will capture all the variables involved in a self-contained manner.

\begin{rem}\label{rem:initial-data-variables} Matrices $\widehat X$, $X$ and $U$ can be initialized with any real matrix of the appropriate dimensions.  When $k\leq T$, all the columns of $\widehat X$, $X$ and $U$ may not be related to the plant states and input, respectively. Our analysis covers this situation and we will be able to guarantee stability properties as desired under appropriate conditions, despite  arbitrary initializations of $\widehat X$, $X$ and $U$. Nevertheless, in practice, it is  reasonable to first run an experiment over a ``physical'' time interval of length $T$ on the system, using possibly an open-loop sequence of inputs, just before $(0,0)$ so that $\widehat X$, $X$ and $U$ are initialized with relevant data at the initial time. \hfill $\Box$
\end{rem}

\subsubsection{Controller variables}\label{subsubsect:pb-statement-controller-variables}
In this work we restrict our attention to the policy class of linear state-feedback controllers, i.e., $u=Kx$, where $K\in\R^{n_u\times n_x}$ is the controller gain to be determined. Because this controller gain will vary at each episode (only), it is natural to model it as a state variable, whose dynamics depend on the most recently collected data, namely $\widehat X$, $X$ and $U$ in (\ref{eq:sys-data-C})-(\ref{eq:sys-data-D}). We also introduce the associated Lyapunov-like matrix $S\in\mathcal{S}_{S}\subset\mathbb{S}_{\succ 0}^{n_x}$ with $\mathcal{S}_S$ bounded, as well as other related auxiliary variables described later $a_1\in[0,1]$, $a_2\in\Rlo$ and $F\in\mathbb{S}^{n_x}_{\succ 0}$. 
As these variables are only updated at each new episode, the dynamics of $K,S,F,a_1,a_2$ when no episode is triggered is,
\begin{eqn}\label{eq:sys-controller-C}
(K^{+},S^{+},F^+,a_1^{+},a_2^+)  & = & 
(K,S,F,,a_1,a_2).
\end{eqn}
On the other hand, when a new episode is triggered $K,S,F, a_1$ and $a_2$ are updated according to
\begin{eqn}\label{eq:sys-controller-D}
(K^{+},S^{+},F^+,a_1^{+},a_2^+) & \in &  L(\widehat X,X,U),
\end{eqn}
where $L:\R^{n_x\times T}\times\R^{n_x\times T}\times\R^{n_u\times T}\rightrightarrows\R^{n_u\times n_x}\times\mathcal{S}_{S}\times\mathbb{S}^{n_x}_{\succ 0}\times [0,1]\times\Rlo$ is a set-valued map to be designed. 
Notice that $\widehat X, X,U$ are only used at the episode triggering instant to define the controller gain $K$ (and the associated variables $S$, $F$,  $a_1$, $a_2$).

\begin{rem}Since $u=Kx$, (\ref{eq:plant-timer-C}) reads $
x^+  =  A(\kappa)x + B(\kappa)K x$,
and the update of $U$ in (\ref{eq:sys-data-C}) between two successive episode-triggering instants is given by
$U^{+}  =  [U_{2:T}, K x]$. This will be reflected in the definition of map $f$ for system (\ref{eq:sys-background}), see  (\ref{eq:sys-f-g}) below. \hfill $\Box$
\end{rem}

\subsubsection{Episode-triggering variables}
We will finally  need some auxiliary variables to design the episode-triggering condition, which we denote by $\eta\in \mathcal{S}_{\eta}\subset\R^{n_\eta}$ with $n_\eta\in\Z_{\geq 1}$. We write the dynamics of the $\eta$-system as, when no episode is being triggered,
\begin{eqn}\label{eq:sys-eta-C}
\eta^{+} & = & h(x,\kappa,\eta)
\end{eqn}
and when an episode is triggered,
\begin{eqn}\label{eq:sys-eta-D}
\eta^{+} & = & \ell(x,\kappa,\eta).
\end{eqn}
Adding extra variables to define triggering conditions is common in the event-triggered control literature, see, e.g., \cite{Girard-tac15,maass-et-al-tac2023(etc)}. 

\subsection{Hybrid model}\label{subsect:hybrid-modeling}

We collect the variables introduced so far to form the state vector
\begin{equation}\label{eq:q}
q:=(x,\kappa,\widehat X,X,U,K,S,F,a_1,a_2,\eta)\in\mathcal{S}_q
\end{equation}
where  
$\mathcal{S}_q :=\R^{n_x}\times\Zo\times\R^{n_x\times T}\times\R^{n_x\times T}\times\R^{n_u\times T}\times
\R^{n_u\times n_x}\times\mathcal{S}_{S}\times \mathbb{S}^{n_x}_{\succ 0}\times [0,1]\times\Rlo\times\mathcal{S}_{\eta}$. 
We can then model the overall closed-loop system as (\ref{eq:sys-background}) with, in view of Section \ref{subsect:additional variables}, 
\begin{eqn}\label{eq:sys-f-g}
{\small\begin{aligned}
f(q)\!:=\!\left(\begin{matrix}A(\kappa)x+B(\kappa)Kx \\ 
\kappa+1 \\
[\widehat{X}_{2:T},x] \\
[X_{2:T}, A(\kappa)x+B(\kappa)Kx]\\
[U_{2:T},K x] \\
K \\
S \\
F \\
a_1 \\
a_2 \\
h(q)\end{matrix}\right),\, 
G(q) \!:=\!  \left(\begin{matrix} x \\ \kappa \\  \widehat X \\ X \\ U \\ L(\widehat X,X,U)\\ \ell(q)\end{matrix}\right),
\end{aligned}}
\end{eqn}
where $f$ is defined over $\mathcal{C}$ and $g$ over $\mathcal{D}$, both sets later defined in Section \ref{subsect:design-episode-triggering}. 

\begin{rem}\label{rem:matrix-valued} Variables $\widehat X,X,U,K,S,F$ are matrix-valued  while $q$ in Section \ref{sect:background} is given as a vector. This inconsistency can easily be overcome by vectorizing these matrix variables. We have  chosen not to do so in (\ref{eq:sys-f-g}) to not over complicate the notation.\hfill $\Box$ 
\end{rem}

\begin{rem}\label{rem:counters}
The reason why we have introduced $\kappa$ is to obtain an autonomous hybrid model of the form of (\ref{eq:sys-background}), like in \cite[Example 3.3]{Goebel-Sanfelice-Teel-book} and \cite{sif-et-al-aut2023}. When a solution $q$ to (\ref{eq:sys-background}) with (\ref{eq:sys-f-g}) is such that $\kappa(0,0)=0$, then $\kappa(k,j)=k$ for any $k\in\Zo$ such that there exists $j\in\Zo$ with $(k,j)\in\dom q$, in other words $\kappa$ corresponds to the physical time $k$. On the other hand, by initializing $\kappa$ to an integer value different from zero, we allow the initial physical time on which $A$ and $B$ depend in  the original plant equation (\ref{eq:plant}) to be non-zero, which is important when dealing with such time-varying systems. In that way, while any solution to $(\ref{eq:sys-background})$ has for initial time $(0,0)$ by Definition \ref{def:solution}, the matrices $A$ and $B$ are allowed to initially depend on $\kappa(0,0)\neq 0$. 
\hfill $\Box$ 
\end{rem}

\subsection{Design objective}\label{subsect:design-objectives}
Consider system (\ref{eq:sys-background}) with (\ref{eq:sys-f-g}), the objective is to design set-valued map $L$, the triggering condition, i.e., the sets  $\mathcal{C}$ and $\mathcal{D}$ as well as $\eta$ and its dynamics, to ensure stability properties for set $\mathcal{A}$ defined as
\begin{equation}\label{eq:set-A-H}
\mathcal{A}:=\left\{q\in\mathcal{S}_q\,:\,x=0\right\}.
\end{equation}
With the notation of Section \ref{subsect:stability-background}, this corresponds to set $\mathcal{A}$ in (\ref{eq:set-A}) with 
\begin{equation}
\begin{aligned}\label{eq:set-A-H-notation-background}
q_1&=x,\quad q_2=(\kappa,\widehat X,X,U,K,S,F,a_1,a_2,\eta),\quad \mathcal{S}_1=\R^{n_x},\\
\mathcal{S}_2&=\Zo\times\R^{n_x\times T}\times\R^{n_x\times T}\times\R^{n_u\times T}\times\R^{n_u\times n_x}\\
&\,\,\,\,\,\times \mathcal{S}_{S}\times\mathbb{S}^{n_x}_{\succ 0}\times [0,1]\times\Rlo\times\mathcal{S}_{\eta}.
\end{aligned}
\end{equation}

We provide for this purpose conditions under which  (P1)-(P3) in Section \ref{sect:background} hold. In particular, in Section \ref{sect:controller-design} we give conditions allowing us to design $L$ and thus the adaptation law for the feedback gain $K$ such that desirable Lyapunov properties are guaranteed on $\mathcal{C}$. In Section \ref{sect:episode-triggering}, we derive Lyapunov properties for system (\ref{eq:sys-background}), (\ref{eq:sys-f-g})  on $\mathcal{D}$. We then merge the results of Sections \ref{sect:controller-design} and \ref{sect:episode-triggering} and show that the established Lyapunov properties imply the satisfaction of (P1)-(P3), and finally derive conditions allowing the application of Theorem \ref{theorem:stability} in Section \ref{sect:stability-guarantees}.

\section{Controller design}\label{sect:controller-design}

We present the method used to design the set-valued map $L$ in (\ref{eq:sys-controller-D}) when an episode is triggered.  

\subsection{Data-based representation }\label{subsect:initial-learning:data-based-modeling}
For any $\kappa\in\Zo$, we define 
\begin{equation}
    \begin{array}{rllll}
    \mathcal{A}(\kappa) & := & \left[A(\kappa-T)\,\ldots\, A(\kappa -1)\right]\in\R^{n_x\times n_x T}\\
    \mathcal{B}(\kappa) & := & \left[B(\kappa -T)\,\ldots\, B(\kappa-1)\right]\in\R^{n_x\times n_u T}.
    \end{array}\label{eq:cal-A-cal-B-initial}
\end{equation}
These  matrices are \emph{not} used for design but only for analysis purpose. The next result is a direct consequence of Lemma \ref{lem:data-based-matrices} and (\ref{eq:cal-A-cal-B-initial}), its proof is therefore omitted.%
\begin{lem}\label{lem:consistent-data-matrices} For any solution $q$ to (\ref{eq:sys-background}), (\ref{eq:sys-f-g}) and  any $(k,j)\in\dom q$ with $k\geq T$, 

\begin{eqn}\label{eq:X0+}
X(k,j)  =  \mathcal{A}(\kappa(k,j))\mathcal{N}(\widehat X(k,j))+\mathcal{B}(\kappa(k,j))\mathcal{N}(\mathcal{U}(k,j)),
\end{eqn}
where the map $\mathcal{N}$ is defined in the notation part. \hfill $\Box$
\end{lem}

To shorten the notation only, we denote the concatenation of the data-matrices $\widehat X,X,U$, with  the counter variable $\kappa$ by 
\begin{eqn}\label{eq:data}
\data:=(\kappa,\widehat X,X,U)\in \mathcal{S}_{\data},
\end{eqn}
with $\mathcal{S}_{\data}:=\Zo\times\R^{n_x\times T}\times\R^{n_x\times T}\times \R^{n_u\times T}$; to help remember its meaning, $\data$ can be associated with the word ``data'' and we will sometimes refer to $\data$ as the data with some slight abuse of terminology. 
For any $\data\in\mathcal{S}_{\data}$ and  $\kappa'\in\Zo$, we define
\begin{equation}\label{eq:D0}
D(\data,\kappa'):=A(\kappa') \widehat X-\mathcal{A}(\kappa) \mathcal{N}(\widehat X)+B(\kappa')U-\mathcal{B}(\kappa)\mathcal{N}(U),
\end{equation}
which belongs to $\R^{n_x\times T}$. We can equivalently rewrite (\ref{eq:D0}) as  
\begin{eqn}\label{eq:D0-Delta}
D(\data,\kappa') & = & (A(\kappa')\otimes \mathbf{1}_T^\top-\mathcal{A}(\kappa))\mathcal{N}(\widehat X) \\
& &+(B(\kappa')\otimes \mathbf{1}_T^\top-\mathcal{B}(\kappa))\mathcal{N}(U). 
\end{eqn}
We can interpret $D(\data,\kappa')$ as a matrix ``measure'' of the distance between the matrices $A(\kappa')$ and $B(\kappa')$ evaluated at a given $\kappa'\in\Zo$, typically greater than or equal to $\kappa$, and the corresponding matrices in $\mathcal{A}(\kappa)$ and $\mathcal{B}(\kappa)$ in (\ref{eq:cal-A-cal-B-initial}) that generated $\data$. 
Equation (\ref{eq:D0-Delta}) shows indeed that, if system (\ref{eq:plant-timer-C}) is time-invariant, then $D(\data,\kappa') \equiv 0$ for any $\kappa'\in\Zo$; otherwise, there exists $\kappa' \in\Zo$ such that $D(\data,\kappa')$ might be non-zero compatibly with the data collected in $\mathcal{N}(\widehat X)$ and $\mathcal{N}(U)$.

\subsection{Matrix proximity sets for LTV systems}\label{subsect:matrix-ellipsoids}

Because gain $K$ is designed based on the values of $\widehat X$, $X$ and $U$ 
at the last episode-triggering instant in view of (\ref{eq:sys-f-g}) and $A$ and $B$ are time-varying, these data  only carry partial information on the true system matrices at the current time. We  
characterize here the uncertainty associated with using the data-based representation (\ref{eq:X0+}) given $\widehat X$, $X$ and $U$ in place of the true plant model (\ref{eq:plant-timer-C}).  

Motivated by the interpretation of $D$ above, we define  the following matrix function for any data  $\data\in\mathcal{S}_{\data}$, $M_A\in\R^{n_x\times n_x}$ and  $M_B\in\R^{n_x\times n_u}$,
\begin{equation}\label{eq:D_tilde}
\widetilde{D}(\data,M_A,M_B):=M_A \widehat X-\mathcal{A}(\kappa) \mathcal{N}(\widehat X)+M_B U-\mathcal{B}(\kappa)\mathcal{N}(U). 
\end{equation}
Clearly $\widetilde{D}(\data,M_A,M_B)=D(\data,\kappa')$ when $M_A=A(\kappa')$ and $M_B=B(\kappa')$ for some $\kappa'\in\Zo$ in view of (\ref{eq:D0-Delta}). 
Using (\ref{eq:D_tilde}), for any $\data\in\mathcal{S}_{\data}$ and $F \in \mathbb{S}^{n_x}_{\succ 0}$, we introduce  the matrix set
\begin{equation}\label{eq:set-E}
\begin{aligned}
\mathcal{E}(\data,F):=\big\{&\left[M_A \, \, M_B
\right]^\top \in \R^{(n_x+n_u)\times n_x} \,:\,\\ 
&\widetilde{D}(\data,M_A,M_B) \widetilde{D}(\data,M_A,M_B)^\top \preceq F\big\},
\end{aligned}
\end{equation}
defining the set of matrices that are \emph{close} to the data-based representation (\ref{eq:X0+}). 
It is useful to introduce this set because, given $\kappa' \in \Zo$, the closeness condition for the system matrices $\left[A(\kappa') \, \, B(\kappa') \right]^\top \in \mathcal{E}(\data,F)$ is equivalent to $D(\data,\kappa')D(\data,\kappa')^\top\preceq F$, which is a condition that can be enforced using robust control tools. We will sometimes omit in the remainder the arguments of $\mathcal{E}$, and of related sets, whenever this is clear from the context.

The definition of set $\mathcal{E}$ raises two important questions. First, despite the quadratic dependence on $M_A$ and $M_B$, it is unclear by simple inspection of (\ref{eq:D_tilde}) under which conditions $\mathcal{E}$ is non-empty and whether it is an ellipsoidal set akin to those encountered in the recent literature on data-driven control of LTI systems \cite{bisoffi-et-al-aut22,vanWaarde_SICO23_QMI}. Second, the set in (\ref{eq:set-E}) depends on  $\mathcal{A}(\kappa)$ and $\mathcal{B}(\kappa)$ and thus on the unknown time-dependent matrix-valued maps  $A(\cdot)$ and $B(\cdot)$ in view of (\ref{eq:D_tilde}),  and therefore cannot be explicitly computed. 
The next result addresses both questions and is an application of recent developments on quadratic matrix inequalities \cite{vanWaarde_SICO23_QMI}.
\begin{lem}\label{lem:ellipsoidal set} 
Given $T\in\Z_{\geq 1}$, let data $\data\in\mathcal{S}_{\data}$ be such that 
\begin{equation}\label{eq:data_consistency}
X=\mathcal{A}(\kappa)\mathcal{N}(\widehat X)+\mathcal{B}(\kappa)\mathcal{N}(U)
\end{equation}
with $\mathcal{A},\mathcal{B}$ in (\ref{eq:cal-A-cal-B-initial}). Define $Z:=\left[\begin{smallmatrix}
        \widehat X \\ U
    \end{smallmatrix}\right]$ and $M  :=Z Z^\top$. Given  $F \in \mathbb{S}^{n_x}_{\succ 0}$ and the associated set $\mathcal{E}(\data,F)$ in (\ref{eq:set-E}), then:
\begin{enumerate}
\item[(i)] $\mathcal{E}(\data,F)$ is non-empty if and only if  
\begin{eqn}
\Delta:=XZ^\top M^{\dagger}Z X^\top-X X^\top+F \succeq 0,
\end{eqn}
\item[(ii)] for any $M_A\in\R^{n_x\times n_x}$ and $M_B\in\R^{n_x\times n_u}$,
\begin{equation}\label{eq:ellips_AB}
\left[ M_A \, \, M_B
\right]^\top\in \mathcal{E}(\data,F) 
\Leftrightarrow 
\left[ M_A \, \, M_B
\right]^\top\in \widehat{\mathcal{E}}(\data,F),
\end{equation}
where $\widehat{\mathcal{E}}(\data,F)$ has the data-based representation
\begin{equation}\label{eq:defEllips}
\begin{aligned}
\widehat{\mathcal{E}}(\data,F) :=\big\{\widehat Z : (\widehat Z-Z_{c})^\top M (\widehat Z-Z_{c}) \preceq \Delta \big\}
\end{aligned}
\end{equation}
with $Z_{c}:=M^{\dagger}
Z X^\top  \in \R^{(n_x+n_u)\times n_x}$. 
\item[(iii)]  $\mathcal{E}(\data,F)$ is bounded if and only if $M \succ 0$.\hfill $\Box$
\end{enumerate}

\end{lem}
We give a concise proof of this result to point out which observations are needed to leverage results from \cite{vanWaarde_SICO23_QMI}.\\
\noindent\textbf{Proof:} 
Let $M_A$ and $M_B$ be of compatible dimensions, we have by (\ref{eq:D_tilde}) and (\ref{eq:data_consistency})
\begin{equation}
\begin{array}{rllll}
\widetilde{D}(\data,M_A,M_B) & = & \left[M_A\,\, M_B\right]Z-X.
\end{array}
\end{equation}
which is a data-based representation of (\ref{eq:D_tilde}). Therefore
\begin{eqn}
& \widetilde{D}(\data,M_A,M_B) \widetilde{D}(\data,M_A,M_B)^\top \preceq F   \\
\Leftrightarrow & \left(\left[M_A\,\, M_B\right]Z-X\right)\left( Z^\top\left[M_A\,\, M_B\right]^\top -X^\top\right) \preceq F\\
\Leftrightarrow &\left[M_A\,\, M_B\right] M \left[M_A\,\, M_B\right]^\top  -\left[M_A\,\, M_B\right]Z X^\top \\
& \hfill -X Z^\top\left[M_A\,\, M_B\right]^\top + X X^\top \preceq F.\\
\Leftrightarrow&\begin{bmatrix}
       I \\  M_A^\top \\  M_B^\top
    \end{bmatrix}^\top
\left[\begin{matrix}
      F-X X^\top & X Z^\top \\  
     Z X^\top & -M \\  
       \end{matrix}\right]
\begin{bmatrix}
      I \\  M_A^\top \\  M_B^\top 
  \end{bmatrix} \succeq 0\\
\end{eqn}
which shows that $\mathcal{E}(\data,F)$ is a set defined by a quadratic matrix inequality with a data based representation. 
Note first that $M \succeq 0$ and the kernel of $M$ is contained or equivalent to the kernel of $X Z^\top$. This allows us to apply the arguments in \cite[Eq. 3.4]{vanWaarde_SICO23_QMI} and in \cite[Eq. 3.3]{vanWaarde_SICO23_QMI} to show item (i) and item (ii), respectively. Item (iii) follows from \cite[Thm. 3.2(b)]{vanWaarde_SICO23_QMI}.

To characterize the shape of the set when $Z$ has not full row rank and thus $M$ is singular, define $m\in[1,n_x+n_u-1]$ the dimension\footnote{The case $m=0$ is not of interest since it corresponds to the system at equilibrium $x=0$.} of the image of $Z$, and $V^\top \in\R^{m\times n_x+n_u}$ and $W^\top \in\R^{(n_x+n_u-m)\times n_x+n_u}$ the matrices built with a basis of the image and the left kernel of $Z$, respectively.
Then it follows from item (ii) that any element $\mathcal{E}(\data,F)$ can be written as
\begin{equation}\label{eq:defEllips_lowrank}
\begin{aligned}
\left[ M_A \, \, M_B
\right]^\top&=
P W^\top + Q V^\top,\quad P \in \R^{(n_x+n_u) \times (n_x+n_u-m)} \\
Q&\in\big\{ \widetilde{Q} \,: \, (\widetilde{Q}^\top-\widetilde Z_{c})^\top \widetilde{M} (\widetilde{Q}^\top-\widetilde Z_{c}) \preceq \widetilde \Delta \big\}\\
\end{aligned}
\end{equation}
where $\widetilde{M}  :=V^\top M  V$, $\widetilde Z_{c}:=-\widetilde M^{-1}\widetilde N$, $\widetilde{N}  := V^\top N$ and $\widetilde{\Delta}:= \widetilde{N}^\top \widetilde{M}^{-1} \widetilde{N}-X X^\top+F$. 

\hfill $\blacksquare$

If $Z$ has full row rank, set $\mathcal{E}$ from (\ref{eq:set-E}) is an ellipsoid in the space of real system matrices of dimensions $(n_x+n_u)\times n_x$ and $\widehat{\mathcal{E}}$ in (\ref{eq:defEllips}) is an equivalent data-based description. When $Z$ is rank deficient, $\widehat{\mathcal{E}}$ has a geometric characterization as an unbounded ellipsoidal-type set (\ref{eq:defEllips_lowrank}). 

\subsection{Lyapunov property}\label{subsect:property}
We exploit set $\mathcal{E}$ in (\ref{eq:set-E}) to define the next central closed-loop property guiding the control and episode-triggering condition design.
\begin{prop}\label{property-lyap-ineq-initial-learning}
Given $T\in\Z_{\geq 1}$ and $\data\in\mathcal{S}_{\data}$, there exist  $a_1\in[0,1]$, $a_2\in\Rlo$, $F\in\mathbb{S}^{n_x}_{\succ0}$, $S\in\mathcal{S}_{S}$ and $K\in \R^{n_u\times n_x}$ such that for any $\varepsilon\in\Rlo$,
\begin{eqn}\label{eq:property-lyap-ineq-initial-learning}\tag{$\mathcal{P}_T(\data)$}
\begin{aligned}
&\left[M_A \, \, M_B
\right]^\top\in \mathcal{E}(\data,F+\varepsilon S^{-1}) 
\Rightarrow \\
&V((M_A+M_B K)x,S) \leq (a_1 +a_2\varepsilon)V(x,S) \quad\forall x\in\R^{n_x},
\end{aligned}
\end{eqn}
with $V(x,S):=x^{\top}S x$ and $\mathcal{E}$ defined in (\ref{eq:set-E}). \hfill $\Box$
\end{prop}
Property \ref{property-lyap-ineq-initial-learning} implies the existence of a piecewise quadratic function $V$ that characterizes the closed-loop response of \emph{any} plant model described by state and input matrices included in the set $\mathcal{E}(\data,F+\varepsilon S^{-1})$. 
This property is used to define the variables $S,F,a_1,a_2$ mentioned in Section \ref{subsubsect:pb-statement-controller-variables}. We see that matrix $S$ plays the role of Lyapunov-like matrix as already hinted at.  
Matrix $F$ characterizes the set of matrices $\mathcal{E}(\data,F)$ for which the function $V$ is guaranteed to decay with rate\footnote{We call $a_1$ decay rate although it can be equal to $1$ with some slight abuse as $a_1$ will typically belong to $[0,1)$.} $a_1$  along the corresponding solutions. On the other hand, matrix $\varepsilon S^{-1}$ allows enlarging the set for which the right hand-side of (\ref{eq:property-lyap-ineq-initial-learning}) holds from $\mathcal{E}(\data,F)$ to $\mathcal{E}(\data,F+\varepsilon S^{-1})$, which may result in the loss of the non-increasing property of $V$ along the corresponding solutions. Indeed, we see that when $\varepsilon\geq 0$ is big enough, $a_1+a_2\varepsilon$ becomes bigger than $1$. Note finally that the choice to take $\varepsilon S^{-1}$ to inflate set $\mathcal{E}(\data,F)$ in place of a generic $E\in\mathbb{S}_{\succeq 0}^{n_x}$  is not restrictive as, given any $E\in \mathbb{S}^{n_x}_{\succeq 0}$, we can always take $\varepsilon:=\min\left\{\varepsilon'\in\Rlo\,:\, \varepsilon' S^{-1} \succeq E\right\}$, which exists as $E \in \mathbb{S}^{n_x}_{\succeq 0}$ and $S \in \mathbb{S}^{n_x}_{\succ 0}$.

\subsection{Intra-episodic control design}\label{subsect:propertyDesign}

Given Property \ref{property-lyap-ineq-initial-learning}, we can define the map $L$ in (\ref{eq:sys-controller-D}) as any set-valued map from $\R^{n_x\times T}\times\R^{n_x\times T}\times\R^{n_u\times T}$ to $\R^{n_u\times n_x}\times\mathcal{S}_{S}\times\mathbb{S}^{n_x}_{\succ 0}\times [0,1]\times\Rlo$ such that, for any $\data\in\mathcal{S}_{\data}$,
\begin{eqn}\label{eq:L}
L(\widehat X,X,U)  \!\subset\!  \Big\{(K',S',F',a_1',a_2') \,  : (\mathcal{P}_T(\data)) \text{ holds}\Big\}.
\end{eqn}
We present below a possible construction of $L$. 
The key observation is that, due to the set inclusion nature of Property \ref{property-lyap-ineq-initial-learning}, $K$ should 
possess some inherent robustness. 
We therefore take inspiration from the results in \cite[Theorem 5]{de-persis-tesi-tac2020} valid for noisy LTI systems, and extend them to our setting. We emphasize that Property \ref{property-lyap-ineq-initial-learning} can also be achieved via alternative robust data-based approaches, e.g., \change{S-lemma \cite{vanWaarde_SICO23_QMI} or Petersen's lemma \cite{bisoffi-et-al-aut22}.}
\begin{proposition}\label{prop:initial-learning-young} 
Given $T\in\Z_{\geq 1}$ consider any $\data\in\mathcal{S}_{\data}$ such that 
\begin{eqn}\label{eq:lem-data-matching-equation}
X &  = & \mathcal{A}(\kappa)\mathcal{N}(\widehat X)+\mathcal{B}(\kappa)\mathcal{N}(U).
\end{eqn}
If there exist $\varsigma \in \Rlp$, $Y\in\R^{T\times n_x},H\in \mathbb{S}^{n_x}_{\succ 0}$ such that the following LMI has a feasible solution
    \begin{equation}\label{eq:KDesign}
    \left[\begin{matrix}
    \widehat{X} Y - \varsigma X X^\top - H & X Y \\
    \star &  \widehat{X} Y
    \end{matrix}\right]  \succ  0,\quad
    \left[\begin{matrix}
    I_T  & Y \\
    \star &  \widehat{X} Y
    \end{matrix}\right]  \succ  0, 
    \end{equation}
then Property \ref{property-lyap-ineq-initial-learning} holds with any $F\in\mathbb{S}^{n_x}_{\succ 0}$ satisfying $
F \prec \frac{\varsigma}{1+\varsigma}H$, 
$a_1=1-a$, $a=\max\big\{a'\in\Rlo\,:\, a' S^{-1} \preceq H-(1+\varsigma^{-1}) F \big\}$,  $a_2=1+\varsigma^{-1}$, $K:=UY(\widehat{X} Y)^{-1}$ and $S= (\widehat{X} Y)^{-1} \succ 0$.
\hfill $\Box$
\end{proposition}

The proof of Proposition \ref{prop:initial-learning-young} is given in the appendix.  
Proposition \ref{prop:initial-learning-young} provides data-based LMI conditions that can be used on-line to obtain variables satisfying Property \ref{property-lyap-ineq-initial-learning}. Contrary to standard results in the recent data-based literature \cite{de-persis-tesi-tac2020,bisoffi-et-al-aut22}, Proposition \ref{prop:initial-learning-young} does not assume persistence of excitation of the collected data sequence, more precisely that $\big[\begin{matrix}
        U^\top \; \widehat{X}^\top
    \end{matrix}\big]^\top$ 
has full row rank. This is important due to the on-line nature of the algorithm, whereby it may be difficult to guarantee a priori that such a condition will be satisfied without adding exploratory signals. The drawback is that the search for the gain $K$ is restricted, in the case of low rank data, to a smaller space as shown in the proof of Proposition \ref{prop:initial-learning-young} in the appendix (cf. (\ref{eq:prop-relation-K-U0-X0})). 
Note that, even if persistence of excitation was verified, the results for noisy data in \cite{de-persis-tesi-tac2020} only give sufficient conditions. 
To reduce this source of conservatism, it is interesting to consider formulations based on the S-lemma \cite{vanWaarde_SICO23_QMI} that, in the noisy LTI case, yield \change{instead} necessary and sufficient conditions; \change{we see this as a meaningful next step and plan to do so in future work.}

In view of Proposition \ref{prop:initial-learning-young}, a possible definition of  $L$ in (\ref{eq:L}) is, for any $\widehat X,X\in\R^{n_x\times T}$ and  $U\in\R^{n_u\times T}$,
\begin{eqn}\label{eq:L-example}
\begin{aligned}
&L(\widehat X,X,U):=\Big\{(K,S,F,a_1,a_2)\,  : \\
& \exists (\varsigma, Y,H)\in\Rlp\times\R^{T\times n_x}\times \mathbb{S}^{n_x}_{\succ 0} \text{ s.t. } (\ref{eq:KDesign}) \text{ holds}, & \\
& K,S,F,a_1,a_2 \text{ as in Proposition \ref{prop:initial-learning-young}}\Big\}.
\end{aligned}
\end{eqn}
This definition allows one to pick any selection of $L$, which can be obtained e.g., by adding an objective function to (\ref{eq:KDesign}).
\begin{rem}\label{rem_choice_T}
The value of $T$ defines the number of columns in the data matrices $\widehat X,X,U$ and in data-driven control works is typically chosen large enough to guarantee that $\big[\begin{matrix}
        U^\top \; \widehat{X}^\top
    \end{matrix}\big]^\top$ 
has full row rank. Because this property is not used here, there is no strict lower bound on its value for our statements to hold. However, the choice of $T$ has an important effect on the map $L$ (\ref{eq:L-example}) and navigates the trade-off between using more of the past information to design the new controller and increasing the set of matrices $\mathcal{E}(\data,F)$ with guaranteed decaying $V$ along the solutions. We currently see this as a tuning parameter and quantifying its impact is an important topic of future research.\hfill $\Box$
\end{rem}


\section{episode-triggering}\label{sect:episode-triggering}

We explain in this section how to define the episode-triggering condition, i.e., how to design sets $\mathcal{C}$ and $\mathcal{D}$ as well as $\eta$ and its dynamics for system (\ref{eq:sys-background}), (\ref{eq:sys-f-g}). 

\subsection{Main idea}\label{subsection:main-idea}

Assuming Property \ref{property-lyap-ineq-initial-learning} holds, the idea is to monitor the Lyapunov function $V$   along the on-going solution. As long as this function  strictly decreases with a certain decay rate there is no need to trigger a new episode as this means a desirable stability property holds. If we detect that this is not the case, we update the controller gain and a new episode is triggered, but this can only occur when $L(\widehat X,X,U)$ is non-empty, which will be reflected in the definition of set $\mathcal{C}$. Similar triggering techniques have been developed in different contexts, see e.g.,  \cite{Mazo-Anta-Tabuada-Aut10,Wang-Lemmon-aut11,maass-et-al-tac2023(etc),de-persis-et-al-tac2023(etc-data)}. 

To formalize this idea, we need to introduce a couple of auxiliary variables, which will form the vector $\eta$ in Section \ref{subsect:hybrid-modeling}. We first introduce $\hat x$, which is essentially the value of $x$ at the last physical time. Hence, the dynamics of the $\hat x$-system is given by, when no episode is triggered,
\begin{eqn}\label{eq:hat-x-C}
\hat x^{+} &  = & x    
\end{eqn}
and, when an episode is triggered and thus the physical time is ``frozen'',
\begin{eqn}\label{eq:hat-x-D}
\hat x^{+} &  = & \hat x.     
\end{eqn}
Thanks to $\hat x$, we can now compare the value of $V$ at the current plant state, namely $V(x,S)$, with its value at the previous physical time, namely $V(\hat x,S)$. We can thus monitor on-line whether
\begin{eqn}\label{eq:episode-condition}
V(x,S) & \leq & \sigma(a_1) V(\hat x,S),
\end{eqn}
where $\sigma(a_1)\in[0,1]$ is a desired decay rate of function $V$ along the solutions to (\ref{eq:sys-background}), (\ref{eq:sys-f-g}). 
In particular, we would take $\sigma(a_1)\in\big[a_1,1]$ so that $\sigma(a_1)$ is greater than or equal to the nominal decay rate of $V$  in Property \ref{property-lyap-ineq-initial-learning}, namely $a_1$. When $\sigma(a_1)<1$, (\ref{eq:episode-condition}) guarantees the strict decrease of $V$ along the considered solution. On the other hand, when $V(x,S)\geq \sigma(a_1)V(\hat x)$, a new episode is triggered if possible, i.e., if $L(\widehat X,X,U)\neq \emptyset$. 
A similar condition to (\ref{eq:episode-condition}) was proposed in \cite{Eising_CDC22_DDswitched} for detecting the active mode in data-based control of switched linear systems. 
Because the values of $x$ and $\hat x$ do not change at each jump corresponding to a new episode in view of (\ref{eq:sys-f-g}) and (\ref{eq:hat-x-D}), the condition $V(x,S)\geq \sigma(a_1)V(\hat x,S)$ may still hold after triggering a new episode. 
If such a situation occurs, this would lead to a Zeno-like behavior in the sense that infinitely many episodes will be triggered in finite physical time. To avoid this shortcoming, we introduce a toggle variable $\tau\in\Zo$. The dynamics of $\tau$ is
\begin{eqn}\label{eq:tau}
\tau^{+}   =  1 \quad q\in \mathcal{C}, & \tau^{+}=0 \quad q\in \mathcal{D}.     
\end{eqn}
We only allow a new episode to be triggered when $\tau=1$, which enforces that at least one physical time step has elapsed since the last episode-triggering instant. This mechanism is needed to avoid episodes to be triggered infinitely many times at the same physical time, using the same set of data.

Having introduced all the auxiliary variables, we can define 
\begin{eqn}\label{eq:eta}
\eta  :=  (\hat x,\tau)\in\mathcal{S}_{\eta} \text{ with }\mathcal{S}_{\eta}:=\R^{n}\times\{0,1\},
\end{eqn}
whose dynamics is given by (\ref{eq:sys-eta-C}) and (\ref{eq:sys-eta-D}) with maps
\begin{eqn}\label{eq:h-ell}
h(q)  :=  (x,1) \,\, \forall q\in\mathcal{C}, & &
\ell(q)     :=  (\hat x,0) \,\, \forall q\in\mathcal{D}.
\end{eqn}

\begin{figure*}[t!]
\begin{equation}\label{eq:C-D}
\begin{array}{lllll}
\mathcal{H}  :  \left\{\begin{array}{llllll}
q^+  =  \left(A(\kappa)x+B(\kappa)Kx, \kappa+1,[\widehat{X}_{2:T},x], [X_{2:T}, A(\kappa)x+B(\kappa)Kx],[U_{2:T},K x],
K,
S,
F,
a_1,
a_2,
x,
1\right) &  q\in \mathcal{C} \\
q^+  \in   \left(x, \kappa,  \widehat X, X, U, L(\widehat X,X,U), \hat x,0\right) &  q\in \mathcal{D}
\end{array}\right.\\
\text{with } q=(x,\kappa,\widehat X,X,U,K,S,F,a_1,a_2,\eta),\,
L(\widehat X,X,U)  \subset  \left\{(K',S',F',a_1',a_2') \,  : (\mathcal{P}_T(\data)) \text{ holds}\right\} \text{ and}\\
\mathcal{C}  :=  \left\{q: V(x,S) \leq \sigma(a_1)  V(\hat x,S) \lor \tau=0 \lor L(\widehat X,X,U)=\emptyset\right\}\\
\mathcal{D}  :=  \left\{q: V(x,S) \geq \sigma(a_1) V(\hat x,S) \land \tau=1\land L(\widehat X,X,U)\neq \emptyset\right\}.
\end{array}
\vspace{-0.8cm}
\end{equation}
\end{figure*}

\subsection{Sets $C$ and $D$ and overall model}\label{subsect:design-episode-triggering}

We define the sets $\mathcal{C}$ and $\mathcal{D}$ of system (\ref{eq:sys-background}), (\ref{eq:sys-f-g}), as in (\ref{eq:C-D}),  
where $V(x,S)=x^\top S x$ as in Property \ref{property-lyap-ineq-initial-learning}. These sets capture the information description in Section \ref{subsection:main-idea}. Indeed, set  $\mathcal{C}$ defines the region of the state space where the Lyapunov function $V$ decays as desired or where an episode has just been triggered (i.e., $\tau=0$) or, lastly, where it is not possible to design a new feedback gain (i.e., $L(\widehat X,X,U)=\emptyset$). Set  $\mathcal{D}$ is defined similarly to enforce the triggering of an episode. The overall model is denoted by $\mathcal{H}$ and is defined in (\ref{eq:C-D}). 

The next result establishes the $k$-completeness of any maximal solution to $\mathcal{H}$ thereby ruling out Zeno-like behavior as already observed above. The proof directly follows by application of Proposition \ref{prop:completeness-background} and the definition of system $\mathcal{H}$, and is therefore omitted.

\begin{proposition}\label{prop:maximal-completeness-solutions-H} Any maximal solution to $\mathcal{H}$ is $k$-complete. \hfill $\Box$
\end{proposition}

\begin{rem}\label{rem:triggering-condition} Other triggering conditions could very well be considered. We could for instance enforce a certain number of physical time instants before checking a state-dependent criterion and not just one as in (\ref{eq:C-D}), like in time-regularized event-triggered control, see, e.g., \cite{borgers-2018(time-reg-etc),Abdelrahim-et-al-aut17,tallapragada-chopra-cdc2012}. We could also envision dynamic triggering rules, which only trigger when a static state-dependent criterion is violated for a certain amount of time,  inspired by \cite{Girard-tac15,wang-et-al-cdc2019}. We leave these extensions for  future work. \hfill $\Box$
\end{rem}

\subsection{Lyapunov property}

To conclude this section, we derive properties of function $V$ at jumps due to the triggering of a new episode. This property will be exploited in Section \ref{sect:stability-guarantees} to ensure the satisfaction of (P3) for the overall model. 

\begin{proposition}\label{prop:lyapunov-properties-episode-jump} For any $q\in\mathcal{D}$ and any\footnote{We acknowledge that we write  
$K^+,S^+,F^+,a_1^+,a_2^+$ with some abuse of notation, this is done  only to simplify the exposition.} $g=(x,\kappa,\widehat X,X,U,K^+,S^+,F^+,a_1^+,a_2^+,\ell(q))\in G(q)$, 
\begin{eqn}\label{eq:prop-lyap-episode-jump}
V(x,S^+) & \leq & \nu_d(q_2) V(x,S)
\end{eqn}
with $\nu_d(q_2)\in\Rlo$ verifying 
$S'\preceq\nu_d(q_2) S$ for any $(K',S',F',a_1',a_2')\in L(\widehat X,X,U)$. 
\hfill $\Box$
\end{proposition}


\noindent\textbf{Proof:} Let $q\in \mathcal{D}$ and  $g=(x,\kappa,\widehat X,X,U,K^+,S^+,a^+,\ell(q))\in G(q)$. As $q\in \mathcal{D}$, $L(\widehat X,X,U)$ is non-empty. We have $V(x,S^+)=x^\top (S^+)^\top x \leq \nu_d(q_2) x^{\top}S x=\nu_d(q_2) V(x,S)$;  note that $\nu_d$ is well-defined as $\mathcal{S}_S$ is bounded and $S\in\mathbb{S}^{n_x}_{\succ 0}$. We have obtained the desired result as $q$ and $g$ have been arbitrarily selected. 
\hfill $\blacksquare$

We can now establish stability properties for system $\mathcal{H}$.

\section{Stability guarantees}\label{sect:stability-guarantees}

We exploit the results of Sections \ref{sect:controller-design} and \ref{sect:episode-triggering} to show that (P1)-(P3) in Section \ref{subsect:background-Lyapunov} are satisfied for system $\mathcal{H}$ with set $\mathcal{A}$  in (\ref{eq:set-A-H}). We  then apply Theorem \ref{theorem:stability} to derive general conditions under which  stability properties hold for system $\mathcal{H}$. Afterwards, we focus on case studies for which we can derive more interpretable stability conditions.

\subsection{Ensuring (P1)-(P3) for $\mathcal{H}$}

We introduce for the sake of convenience the next subset of $\mathcal{C}$ as defined in (\ref{eq:C-D})
\begin{eqn}\label{eq:C1}
\mathcal{C}_1  & := & \left\{q\in\mathcal{S}_q \,:\, V(x,S) \leq \sigma(a_1)  V(\hat x,S)\right\}. 
\end{eqn}
We establish in the following result that (P1)-(P3), as stated in Section \ref{subsect:background-Lyapunov}, hold for system $\mathcal{H}$.

\begin{proposition}\label{prop:P1-P3} Given $T\in\Z_{\geq 1}$, 
consider system $\mathcal{H}$. Then  (P1)-(P3) hold with 
\begin{eqn}\label{eq:prop-functions-P1-P3}
\mathcal{U}(q) & := &  V(x,S)  &  \forall q\in\mathcal{S}_q \\
\underline\alpha(s,q_2) & = & \lambda_{\min}(S) s^2 & \forall (s,q_2)\in\Rlo\times\mathcal{S}_{2} \\
\overline\alpha(s,q_2) & = & \lambda_{\max}(S) s^2 & \forall (s,q_2)\in\Rlo\times\mathcal{S}_{2} \\
\nu_c(q) & =  & \left\{\begin{array}{lllll}
\sigma(a_1) & f(q)\in \mathcal{C}_1 \\
\theta(q_2)  & f(q)\notin \mathcal{C}_1 \end{array}\right.        &  \forall q\in \mathcal{C},
\end{eqn}
$\nu_d$ as in Proposition \ref{prop:lyapunov-properties-episode-jump}, $V$ as in Property \ref{property-lyap-ineq-initial-learning},  $\theta(q_2)\in[\underline\theta(q_2),\infty)$, $ \underline\theta(q_2):=
\min\big\{\theta'\in\Rlo\,:\,(A(\kappa)+B(\kappa)K)^\top S(A(\kappa)+B(\kappa)K) \leq \theta' S  \big\} $,  $q_2$ as in (\ref{eq:set-A-H-notation-background}). \hfill $\Box$
\end{proposition}

\noindent\textbf{Proof:} Let $q\in \mathcal{C}\cup\mathcal{D}$, by definition of $\mathcal{U}$ in (\ref{eq:prop-functions-P1-P3}), $\mathcal{U}(q)=V(x,S)=x^\top S x$. 
As $S\in\mathcal{S}_S\subset\mathbb{S}^{n_x}_{\succ 0}$, $\lambda_{\min}(S) |x|^2\leq \mathcal{U}(q)\leq \lambda_{\max}(S) |x|^2$ with $0<\lambda_{\min}(S)\leq\lambda_{\max}(S)$. This proves (P1) holds with $\underline\alpha$ and $\overline\alpha$ in (\ref{eq:prop-functions-P1-P3}).

Let $q\in\mathcal{C}$, $\mathcal{U}(f(q)) = V(A(\kappa)x+B(\kappa)Kx,S)$. 
When $f(q)\in\mathcal{C}_1$, by definition of $f$ in (\ref{eq:sys-f-g}) and of $\mathcal{C}_1$ in (\ref{eq:C1}),  $V(A(\kappa)x+B(\kappa)Kx,S)\leq \sigma(a_1)V(x,S)$. Consequently, 
$\mathcal{U}(f(q))  \leq  \sigma(a_1)\mathcal{U}(q)$. 
On the other hand, when $f(q)\notin\mathcal{C}_1$, by definition of $\theta(q)$,
$\mathcal{U}(f(q)) \leq  \theta(q_2)\mathcal{U}(q)$. 
Note that $\underline\theta(q_2)$ in Proposition \ref{prop:P1-P3} is well-defined as $(A(\kappa)+B(\kappa)K)^\top S(A(\kappa)+B(\kappa)K)\succeq 0$ and $S\succ 0$. We have proved that (P2) holds with $\nu_c$ in (\ref{eq:prop-functions-P1-P3}). 
Finally, the satisfaction of (P3) with $\nu_d$ as in (\ref{eq:prop-functions-P1-P3}) follows by Proposition \ref{prop:lyapunov-properties-episode-jump}. \hfill $\blacksquare$

Note that $\theta(q_2)$  in Proposition \ref{prop:P1-P3} is greater than $\sigma(a_1)$ when $f(q)\notin \mathcal{C}_1$, as $f(q)$ would be in $\mathcal{C}_1$ otherwise. We show below how to relate $\theta$ to parameters of Property \ref{property-lyap-ineq-initial-learning} in Section \ref{subsect:matrix-ellipsoids}. 

\subsection{Bound on $\mathcal{U}$ along  the solutions to $\mathcal{H}$}

The next lemma follows directly from  Proposition \ref{prop:P1-P3}, by application of Lemma \ref{lem:lyap-bound} and noting that $a_1$ is constant on $\mathcal{C}$, see (\ref{eq:sys-f-g}). Its proof is therefore omitted.

\begin{lem}\label{lem:lyap-bound-H}  Given $T\in\Z_{\geq 1}$, consider system $\mathcal{H}$, for any solution $q$ it holds that 
\begin{eqn}\label{eq:lemma-lyap-bound-H}
\mathcal{U}(q(k,j)) & \leq  \pi(q,k,j)\mathcal{U}(q(0,0)) & & \forall (k,j)\in\dom q,
\end{eqn}
with $\mathcal{U}$ as in Lemma \ref{prop:P1-P3}, 
\begin{eqn}\label{eq:pi-H}
{\small\begin{aligned}
&\pi(q,k,j)  = \dst\prod_{j'=0}^{j-1}\underbrace{\prod_{\substack{k'=k_{j'} \\ k'\in\mathcal{T}_{1}(q)}}^{k_{j'+1}-1}\!\!\sigma(a_1(k_{j'},j'))\nu_d(q_2(k_{j'+1},j'))}_{\text{(A)}}\\
&\dst \times \!\!\underbrace{\prod_{\substack{k'=k_{j'} \\ k'\notin\mathcal{T}_{1}(q)}}^{k_{j'+1}-1}\!\!\theta(q_2(k',j'))}_{\text{(B)}} 
\!\underbrace{\prod_{\substack{k''=k_{j} \\ k''\in\mathcal{T}_1(q)}}^{k-1}\sigma(a_1(k_j,j))\!\!\prod_{\substack{k''=k_{j} \\ k''\notin\mathcal{T}_1(q)}}^{k-1}\!\!\!\!\theta(q_2(k'',j))}_{\text{(C)}}
\end{aligned}}
\end{eqn}
$\mathcal{T}_1(q)  := \big\{k\in\Zo\,:\exists j\in\Zo, \,(k,j),(k+1,j)\in\dom q, q(k+1,j)\in  \mathcal{C}_1\big\}$,  $k_0,k_1,\ldots, k_{j}$ as defined at the end of Section \ref{subsection:background-solution},  $\mathcal{C}_1$ in (\ref{eq:C1}) and $\theta$ in Proposition \ref{prop:P1-P3}. \hfill $\Box$ 
\end{lem}

Despite its apparent complexity,  (\ref{eq:lemma-lyap-bound-H})-(\ref{eq:pi-H}) admit an intuitive interpretation. The term (A) is made of the product of the desired decay rate $\sigma(a_1)$ with $\nu_d(q_2)$, which is due to the potential change of value of $S$ at episode triggering, along solutions. The term (B) characterizes the potential growth of $\mathcal{U}$ when the desired decay rate $\sigma(a_1)$ is not met. This allows quantifying, for example, the growth of $\pi$ when a controller update is triggered but $L(\widehat X,X,U)=\emptyset$. Finally, the term (C) is the product of $\sigma(a_1)$ and of the potential growth quantified by $\theta$ when $k\in[k_j,k_{j+1}]$. 

As long as the solution remains in $\mathcal{C}_1$, the growth rate of $\pi(q,\cdot,\cdot)$ is upper-bounded by the desired rate $\sigma(a_1)$ and we will be able to derive that $\pi$ and thus $\mathcal{U}(q)$ converges to the origin along the considered solution (under mild extra conditions). 
The  potential obstacle for the convergence of $\mathcal{U}(q)$ to $0$ as $k+j$ goes to infinity is if the solution wanders too often and for too long periods of physical times outside $\mathcal{C}_1$. This may be due to the fact that the matrices $A$ and $B$ are changing significantly too often, so that after a new episode the solution does not remain in $\mathcal{C}_1$ for long enough periods of physical time, 
or that the lastly collected data $\widehat X$, $X$, $U$ do not yield a new controller gain $K$, i.e., $L(\widehat X,X,U)=\emptyset$. 
Later we will capture analytically these effects by providing sufficient conditions for stabilization that can be related to the system's time-variations. 

Before proceeding further, it is essential to relate $\theta$ in (\ref{eq:pi-H}), which  characterizes the growth of $\mathcal{U}$ when the solutions to $\mathcal{H}$ are not in set $\mathcal{C}_1$, to the parameters of the matrix sets in Property \ref{property-lyap-ineq-initial-learning}. This relation is derived in the next lemma. 


\begin{lem}\label{lem:theta} Given $T\in\Z_{\geq 1}$, consider system $\mathcal{H}$. For any solution $q$ and any $(k,j)\in\dom q$ with $k\geq T$ and $j\geq j_T+1$, there exists $\varepsilon:\mathcal{S}_{\data}\to\Rlo$ such that  
$[A(\kappa(k,j))\,\,B(\kappa(k,j))]^\top\in\mathcal{E}\Big(\data(k,j),F(k,j)+\varepsilon(\data(k,j)) S^{-1}(k,j)\Big)$,
with $\mathcal{E}$ in (\ref{eq:set-E}), $\data$ in (\ref{eq:data}), and Lemma \ref{lem:lyap-bound-H} holds with 
$\theta(q_2(k,j))=a_1(k,j)+a_2(k,j)\varepsilon(\data(k,j))$.  
\hfill $\Box$ 
\end{lem}

\noindent\textbf{Proof:}  Let $q$ be a solution to $\mathcal{H}$ and $(k,j)\in\dom q$ with $k\geq T$ and $j\geq j_{T+1}$. 
By Lemma \ref{lem:consistent-data-matrices} and item \change{(i)} of Lemma \ref{lem:ellipsoidal set}, the set $\mathcal{E}(\data(k,j),F(k,j)+\varepsilon S^{-1}(k,j))$ is non-empty for  $\varepsilon\in\Rlp$ sufficiently big as $S(k,j)\in\mathbb{S}_{\succ 0}^{n_x}$. Furthermore, in view of the definition of $\mathcal{E}$ in (\ref{eq:set-E}), we can always find $\varepsilon(\data(k,j))\in\Rlo$ sufficiently big such that   $\change{[A(\kappa(k,j))\,\,B(\kappa(k,j))]^\top}\in\mathcal{E}\Big(\data(k,j),F(k,j)+\varepsilon(\data(k,j)) S^{-1}(k,j)\Big)$. Consequently, as  $(\mathcal{P}_T(\data(q(k_j,j))))$ holds by definition of $L$ in (\ref{eq:L}),  
\begin{equation}\nonumber
\begin{array}{l}
V((A(\kappa(k,j))+B(\kappa(k,j))K(k,j))x(k,j),S(k,j)) \\ \quad\quad\quad\leq \big(a_1(k,j)+a_2(k,j)\varepsilon(\data(k,j)\big) V(x(k,j),S(k,j)).
\end{array}
\end{equation}
In view of the definition of $\underline\theta$ in Proposition \ref{prop:P1-P3}, $\underline\theta(q_2(k,j))\leq a_1(k,j)+a_2(k,j)\varepsilon(\data(k,j))$, which means we can select $\theta(q_2(k,j))= a_1(k,j)+a_2(k,j)\varepsilon(\data(k,j))$ in Lemma \ref{lem:lyap-bound-H}. \hfill $\blacksquare$

Lemma \ref{lem:theta} quantifies the relation between the distance of the plant matrices $[B(\kappa)\,A(\kappa)]$ from the matrix set $\mathcal{E}(\data,F)$ of guaranteed decay (characterized by $\varepsilon S^{-1}$) and the growth rate $\theta$ of $\mathcal{U}$ along the corresponding solutions. Lemma \ref{lem:theta} also allows us to derive a  more insightful expression for $\pi$ in (\ref{eq:pi-H}) by replacing the terms (B) and (C) as below
{\small\begin{equation}\label{eq:pi-H-bis}
\begin{array}{lllll}
\text{(B)}  =   \!\!\!\!\underbrace{\prod_{\substack{k'=k_{j'} \\ k'\notin\mathcal{T}_{1}(q) \\ j'\leq j_T +1}}^{k_{j'+1}-1}\!\!\!\!\!\!\theta(q_2(k',j'))}_{\text{(B1)}}\!
 \underbrace{\dst\prod_{\substack{k'=k_{j'} \\ k'\notin\mathcal{T}_{1}(q) \\ j'\geq j_T+1}}^{k_{j'+1}-1}\!\!\!\!(a_1(k',j')\!+\!a_2(k',j')\varepsilon(q_2(k',j')))}_{\text{(B2)}}\\
\text{(C)}  = \prod_{\substack{k''=k_{j} \\ k''\in\mathcal{T}_1(q)}}^{k-1}\!\!\sigma(a_1(k_j,j))\\
\quad\quad\times\prod_{\substack{k''=k_{j} \\ k''\notin\mathcal{T}_1(q)}}^{k-1}\!\!\!\!(a_1(k'',j)+a_2(k'',j)\varepsilon(q_2(k'',j))).
\end{array}
\end{equation}}
The term (B1) is due to what happens before the $j_{T}+1^{\text{th}}$ episode has occurred, during which we do not have much control of the Lyapunov-like function $V$ unless we know a stabilizing policy valid for the first physical time steps for instance. This is reasonable when we have a good knowledge of the initial values of $A$ and $B$, but these may significantly vary afterwards thereby justifying the proposed approach. The term (B2) characterizes the potential growth of $\mathcal{U}$ when the desired decay rate $\sigma(a_1)$ is not verified. We see that the corresponding terms depend on $\varepsilon$ as in Lemma \ref{lem:theta}, which characterizes the ``distance'' of $[A(\kappa)\,\,B(\kappa)]^\top$ to the set $\mathcal{E}(\data,F)$ for which decay is guaranteed as discussed in Section \ref{subsect:property}. 
Similar interpretations hold for term (C).




\subsection{General stability conditions}\label{subsect:stability-general}

The next theorem follows by application of  Theorem \ref{theorem:stability}  to system $\mathcal{H}$ with $\pi$ in (\ref{eq:pi-H}).

\begin{thm}\label{thm:stability-H} Given $T\in\Z_{\geq 1}$, consider system $\mathcal{H}$. The following holds. \begin{enumerate}[label=(\roman*)]
\item If there exists $\mu:\Rlo\times\mathcal{S}_2\to\Rlo$ continuous and non-decreasing in its first argument such that for any solution $q$ and any $(k,j)\in\dom q$, 
$\pi(q,k,j)\frac{\lambda_{\max}(S(0,0))}{\lambda_{\min}(S(k,j))}\leq  \mu(|q_1(0,0)|,q_2(0,0))$ 
with $q_1,q_2,\mathcal{S}_2$ in (\ref{eq:set-A-H-notation-background}), $\pi$ in (\ref{eq:pi-H}), then $\mathcal{A}$ is stable. In addition, 
\begin{enumerate}[leftmargin=1cm]
    \item[(i-a)] if $\mu(s,\cdot)$ is constant for any $s\in\Rlo$, then $\mathcal{A}$ is uniformly stable,
    \item[(i-b)] if for any complete solution $q$, $\pi(q,k,j)\lambda_{\min}(S(k,j))^{-1}\to 0$ as $k+j\to\infty$ then $\mathcal{A}$ is GAS.
\end{enumerate}
\item If there exists $\beta\in\mathcal{L}$  such that  for any solution $q$, 
$\pi(q,k,j)\frac{\lambda_{\max}(S(0,0))}{\lambda_{\min}(S(k,j))} \leq  \beta(k+j)$ for any $ (k,j)\in\dom q$, 
then $\mathcal{A}$ is UGAS.
\item If there exist  $c_1\geq 1$, $c_2>0$ such that  for any solution $q$, 
$\pi(q,k,j)\frac{\lambda_{\max}(S(0,0))}{\lambda_{\min}(S(k,j))}  \leq  c_1 e^{-c_2(k+j)}$ for any $ (k,j)\in\dom q$, 
then $\mathcal{A}$ is UGES.
\hfill $\Box$
\end{enumerate}
\end{thm}

\noindent\textbf{Proof:} (i) Let $q$ be a solution to $\mathcal{H}$. By Proposition \ref{prop:P1-P3} and Lemma \ref{lem:lyap-bound-H}, for any $(k,j)\in\dom q$,
\begin{eqn}\label{eq:proof-thm-stability-H}
\begin{aligned}
&|q_1(k,j)|  \leq  \big(\tfrac{\lambda_{\max}(S(0,0))}{\lambda_{\min}(S(k,j))} \pi(q,k,j) \big)^{\tfrac{1}{2}}|q_1(0,0)| \\
&\leq  \mu(|q_1(0,0)|,q_2(0,0)) ^{\tfrac{1}{2}}|q_1(0,0)|=:\vartheta(|q_1(0,0)|,q_2(0,0)).
\end{aligned}
\end{eqn}
The map\footnote{If $\mu\equiv 0$, we can take any $\vartheta:\Rlo\times\mathcal{S}_2\to\Rlo$ that is $\Kinf$ in its first argument and the result holds.}  $\vartheta$ 
is  of class-$\mathcal{K}_\infty$ in its first argument. As $q$ has been arbitrarily chosen,  we derive by item (i) of Theorem \ref{theorem:stability} that $\mathcal{A}$ is stable for system $\mathcal{H}$. Items (i-a) and (i-b) of Theorem \ref{thm:stability-H} follow by application of items (i-a) and (i-b) of Theorem \ref{theorem:stability}, respectively.

(ii) Let $q$ be a solution to $\mathcal{H}$. By 
(\ref{eq:proof-thm-stability-H}), for any $(k,j)\in\dom q$, 
$|q_1(k,j)|  \leq \beta(k+j)^{\tfrac{1}{2}}|q_1(0,0)|$. 
As the map from $\Rlo\times\Rlo$ to $\Rlo$ defined as $(s,t)\mapsto \beta(t)^{\tfrac{1}{2}}s$ is of class-$\mathcal{KL}$ and $q$ has been arbitrarily chosen, we deduce from item (ii) of Theorem \ref{theorem:stability} that $\mathcal{A}$ is UGAS. Item (iii) of Theorem \ref{thm:stability-H} is proved in a similar manner. \hfill $\blacksquare$

A strength of Theorem \ref{thm:stability-H} lies in the generality of the (sufficient) conditions it proposes under which set $\mathcal{A}$ in (\ref{eq:set-A-H}) exhibits stability properties. These conditions implicitly impose restrictions on the matrices $A$ and $B$ and their rate of change, thereby allowing to  cover a range of scenarios in a unified manner. A consequence of this is that the requirements of Theorem \ref{thm:stability-H}  may  be difficult to check as they involve  \emph{any} solution to system $\mathcal{H}$. Nevertheless, given an unknown LTV system, we may exploit its  features (like  e.g., slowly varying or sporadically varying plant matrices) to show that the requirements of Theorem \ref{thm:stability-H} hold. We demonstrate it in the rest of this section where we provide easier-to-check conditions, that can be related to the matrices $A$ and $B$ notably via the matrix set characterization in Section \ref{subsect:matrix-ellipsoids} and Property \ref{property-lyap-ineq-initial-learning}. 
We emphasize that these are only a sample of conditions ensuring satisfaction of Theorem \ref{thm:stability-H}.

\subsection{Case studies}\label{subsect:stability-special-cases}
\subsubsection{Solutions eventually always in $\mathcal{C}_1$}
The next result provides sufficient conditions to derive stability properties for $\mathcal{A}$ for system $\mathcal{H}$ in the case where all solutions eventually always lie in $\mathcal{C}_1$ defined in (\ref{eq:C1}). 

\begin{thm}\label{thm:stability-eventually-in-C1} Given $T\in\Z_{\geq 1}$, consider system $\mathcal{H}$. Suppose there exist $T^\star:\mathcal{S}_2\to\Zo$ and $\mu_1:\mathcal{S}_2\to \Rlo$ such that the following holds for any solution $q$ and $(k,j)\in\dom q$.
\begin{enumerate}[label=(\roman*)]
\item When $k\geq T^\star(q_2(0,0))$, $q(k,j)\in\mathcal{C}_1$.
\item When $k\leq T^{\star}(q_2(0,0))$,  $\pi(q,k,j)\frac{\lambda_{\max}(S(0,0))}{\lambda_{\min}(S(k,j))}\leq \mu_1(q_2(0,0))$.
\end{enumerate}
Then $\mathcal{A}$ is stable. In addition, 
\begin{itemize}
\item if $\mu_1$  is constant, then $\mathcal{A}$ is uniformly stable,
\item if $\sigma(a_1(T^{\star}(q_2(0,0)),j_{T^{\star}}(q_2(0,0))))\in[0,1)$ where $j_{T^\star}(q_2(0,0))=\max\{j\in\Zo\,:\,(T^\star(q_2(0,0)),j)\in\dom q\}$, then $\mathcal{A}$ is GAS,
\item if $\mu_1$ and $T^\star$ are constant and there exists $\mu_2\in(0,1)$ such that $\sigma(a_1(T^\star,j_T^{\star}))\leq \mu_2$ for any solution, then $\mathcal{A}$ is UGES. \hfill $\Box$
\end{itemize}
\end{thm}

\noindent\textbf{Proof:} Let $q$ be a solution to $\mathcal{H}$. We denote for the sake of convenience $(T^{\star}(q_2(0,0)),j_{T^\star}(q_2(0,0))$ by $(T^\star,j_{T^\star})$. 
Let $(k,j)\in\dom q$ with $k\geq T^\star$. We observe that by definition of $\pi$ in (\ref{eq:pi-H}), 
\begin{eqn}
\pi(q,k,j) & \leq  \pi(q,T^\star,j_{T^\star})\pi(\widetilde q,k-T^\star,j-j_{T^\star}),
\end{eqn}
with $\widetilde q(k,j)=q(k+T^{\star},j+j_{T^{\star}})$, which is a solution to $\mathcal{H}$ initialized at $q(T^\star,j_{T^\star})$. Hence, by item (ii) of Theorem \ref{thm:stability-eventually-in-C1}, 
\begin{eqn}
\pi(q,k,j)\frac{\lambda_{\max}(S(0,0))}{\lambda_{\min}(S(k,j))} & \leq  \mu_1(q_2(0,0))\pi(\widetilde q,k-T^\star,j-j_{T^\star}).
\end{eqn}
On the other hand, by item (i) of Theorem \ref{thm:stability-eventually-in-C1} and the definition of $\pi$ in (\ref{eq:pi-H}), we derive that $\pi(\widetilde q,k-T^\star,j-j_{T^\star})=\sigma(a_1(T^\star,j_{T^\star}))^{k-T^\star}$ and thus that 
\begin{eqn}\label{eq:proof-bound-U-H-eventually-in-C1}
\pi(q,k,j)\frac{\lambda_{\max}(S(0,0))}{\lambda_{\min}(S(k,j))} & \leq  \mu_1(q_2(0,0)) \sigma(a_1(T^\star,j_{T^\star}))^{k-T^\star}.
\end{eqn}
As $\sigma(a_1(T^\star,j_{T^\star}))\in[0,1]$ in view of (\ref{eq:episode-condition}),
\begin{eqn}\label{eq:proof-eventually-in-C1-stability-after-T-star}
\pi(q,k,j)\frac{\lambda_{\max}(S(0,0))}{\lambda_{\min}(S(k,j))} & \leq  \mu_1(q_2(0,0)).
\end{eqn}
We deduce from item (ii) of Theorem \ref{thm:stability-eventually-in-C1} and (\ref{eq:proof-eventually-in-C1-stability-after-T-star}) that item (i) of Theorem \ref{thm:stability-H} is verified with $\mu(s_1,s_2)=\mu_1(s_2)$. Consequently, $\mathcal{A}$ is stable. When $\mu_1$ is constant, we have by item (i-a) by Theorem \ref{thm:stability-H} that $\mathcal{A}$ is uniformly stable. 
When $ \sigma(a_1(T^\star,j_{T^\star}))\in[0,1)$, by (\ref{eq:proof-bound-U-H-eventually-in-C1}), $\pi(q,k,j)\lambda_{\min}(S(k,j))^{-1}\to 0$ as $k\to\infty$ when $q$ is  maximal; recall that it is  $k$-complete by Proposition \ref{prop:maximal-completeness-solutions-H}. As a consequence, $\mathcal{A}$ is GAS.

Finally, when $\mu_1$ and $T^\star$ are constant, and $\sigma(a_1(T^\star,j_{T^\star}))\leq \mu_2\in(0,1)$ for any solution, (\ref{eq:proof-bound-U-H-eventually-in-C1})  becomes for any $(k,j)\in\dom q$ with $k\geq T^\star$
\begin{eqn}
\pi(q,k,j)\frac{\lambda_{\max}(S(0,0))}{\lambda_{\min}(S(k,j))}  \leq  \mu_1   \mu_2^{k-T^\star}  \! = \! \mu_1 \mu_2^{-j_{T^\star}-T^\star}\mu_2^{k+j_{T^{\star}}}
\end{eqn}
where we write $\mu_1(q_{2}(0,0))=\mu_1$  with some slight abuse of notation. Noting that $j=j_{T^\star}$  when $k\geq T^{\star}$ by item (i) of Theorem \ref{thm:stability-eventually-in-C1}, we derive
\begin{eqn}\label{eq:proof-eventually-in-C1-KL-after-T-star}
\pi(q,k,j)\frac{\lambda_{\max}(S(0,0))}{\lambda_{\min}(S(k,j))} & \leq \mu_1 \mu_2^{-T^\star-j_{T^\star}}\mu_2^{k+j}.
\end{eqn}
On the other hand, from item (ii) of Theorem \ref{thm:stability-eventually-in-C1}, for any $(k,j)\in\dom q$ with $k\leq T^{\star}$,
\begin{eqn}\label{eq:proof-eventually-in-C1-KL-before-T-star}
\pi(q,k,j)\frac{\lambda_{\max}(S(0,0))}{\lambda_{\min}(S(k,j))} & \leq  \mu_1   \leq \mu_1 \mu_2^{-T^\star-j_{T^\star}} \mu_2^{k+j}.
\end{eqn}
We deduce from (\ref{eq:proof-eventually-in-C1-KL-after-T-star}) and (\ref{eq:proof-eventually-in-C1-KL-before-T-star}) that item (iii) of Theorem \ref{thm:stability-H} holds with $c_1=\mu_1\mu_2^{-T^{\star}-j_{T^{\star}}}$ and $c_2=\ln(\mu_2)$. As a result, $\mathcal{A}$ is UGES. \hfill $\blacksquare$

Item (i) of Theorem \ref{thm:stability-eventually-in-C1} means that all solutions to $\mathcal{H}$ eventually lie for all future hybrid times (recall that hybrid times are  pairs $(k,j)$, see Section \ref{subsection:background-solution}) in set $\mathcal{C}_1$. This typically occurs when matrices $A(\kappa)$ and $B(\kappa)$ eventually stop varying significantly as formalized in the corollary below. 
Item (ii) of Theorem \ref{thm:stability-eventually-in-C1}  implies that we can upper-bound the norm of $q_1$ from physical time $0$ to physical time $T^\star(q_2(0,0))$ (at which $q$ enters forever in $\mathcal{C}_1$) by the product of a function of the initial value of $q_2$ with the initial value of $|q_1|$. 
Then, stricter conditions are derived to ensure stronger stability properties. The next corollary essentially provides conditions on $A$, $B$ and $K$ such that the requirements of Theorem \ref{thm:stability-eventually-in-C1} hold.

\begin{cor}\label{cor:eventually-in-C1} Given $T\in\Z_{\geq 1}$, consider system $\mathcal{H}$ with $\sigma(a_1)\in[a_1,1)$ for any $a_1\in[0,1)$ and $\sigma(1)=1$. Suppose there exist $T^\star:\mathcal{S}_2\to \Z_{\geq T}$ and $\overline m\in\Rlo$, such that the following holds for any solution $q$ and any $(k,j)\in\dom q$. 
\begin{enumerate}[label=(\roman*)]
    \item When $k\geq T^{\star}$ and $j\geq j_T+1$, 
    \begin{equation}\nonumber
[A(\kappa(k,j))\,\,\, B(\kappa(k,j))]^\top\in\mathcal{E}(\data(T^{\star},j_{T^\star}),F(T^{\star},j_{T^\star})),
    \end{equation}
    where we omit the dependence of $T^\star$ on $q_2(0,0)$ and  $j_{T^\star}=\max\{j'\in\Zo\,:\,(T^\star,j')\in\dom q\}$. 
    \item  $\max\{\|A(\kappa(k,j))\|,\|B(\kappa(k,j))\|,\|K(k,j)\|\}\leq \overline m$. 
\end{enumerate}
Then $\mathcal{A}$ is stable. In addition, 
\begin{itemize}
\item if $T^\star$  is constant, then $\mathcal{A}$ is uniformly stable,
\item if $a_1(T^\star,j_{T^\star})\in[0,1)$, then $\mathcal{A}$ is GAS,
\item if  $T^\star$ is constant and there exists $c\in(0,1)$ independent of $q$ such that 
\begin{equation}\label{eq:prop-eventually-in-C1-uniform-bound-a1}
a_1(T^\star,j_{T^\star})\leq  c,
\end{equation}
then $\mathcal{A}$ is UGES. \hfill $\Box$
\end{itemize}
\end{cor}

\noindent\textbf{Proof:} Let $q$ be a solution to $\mathcal{H}$ and $(k,j)\in\dom q$ with $k\geq T^\star$ and $j\geq j_T+1$. 
Since $T^{\star}\geq T$ in  item (i) of Corollary \ref{cor:eventually-in-C1}, by definition of $L$ in (\ref{eq:L}), $(\mathcal{P}_T(\data(T^{\star},j_{T^\star})))$ holds. As a consequence, for any $k > T^\star$, 
$
V(x(k,j_{T^\star}),S(k,j_{T^\star}))  \leq  a_1(T^{\star},j_{T^\star}) V(\hat x(k,j_{T^\star}),S(k,j_{T^\star}))$. 
Since $a_1(T^{\star},j_{T^\star}) \leq \sigma(a_1(T^{\star},j_{T^\star}))$ by definition of $\sigma$ in Corollary \ref{cor:eventually-in-C1}, the above inequality implies that 
$q(k,j_{T^\star})\in \mathcal{C}_1$ for any $k > T^\star$. 
As a consequence, item (i) of Theorem \ref{thm:stability-eventually-in-C1} holds with $T^\star+1$. Item (ii) of  Corollary \ref{cor:eventually-in-C1} implies that $|x(k,j)|\leq (\overline m+\overline m^2)^{T^\star}$ for any $(k,j)\in\dom q$ with $k\leq T^\star$. We can then follow similar lines as in the proof of Theorem \ref{thm:stability-eventually-in-C1} to derive that $\mathcal{A}$ is stable as, although the term $(\overline m+\overline m^2)^{T^\star}$ does not necessarily upper-bounds $\pi(q,k,j)\frac{\lambda_{\max}(S(0,0))}{\lambda_{\min}(S(k,j))}$, it does so for  $|x(k,j)|$ which is needed in view of the proof of Theorem \ref{theorem:stability}. When $T^\star$ is constant, $|x(k,j)|$ is upper-bounded by a constant for $k\leq T^\star$, and we derive that $\mathcal{A}$ is uniformly stable like in Theorem \ref{thm:stability-eventually-in-C1}.  When $a_1(T^{\star},j_{T^\star})\in[0,1)$ for any solution $q$, $\sigma(a_1(T^{\star},j_{T^\star}))\in[0,1)$ and we derive from Theorem \ref{thm:stability-eventually-in-C1} that $\mathcal{A}$ is GAS. Finally, when $T^\star$ is constant and (\ref{eq:prop-eventually-in-C1-uniform-bound-a1}) holds,  we conclude that $\mathcal{A}$ is UGES by the last item of Theorem \ref{thm:stability-eventually-in-C1}. \hfill $\blacksquare$

Item (i) of Corollary \ref{cor:eventually-in-C1} means that the matrices $A(\kappa)$ and $B(\kappa)$ eventually stay in set $\mathcal{E}(\data,F)$, which requires these matrices not to vary significantly after some time. Item (ii) of Corollary \ref{cor:eventually-in-C1} on the other hand simply requires the plant matrices and the controller to be norm-bounded, which is a mild condition.

\subsubsection{Solutions frequently enough in $\mathcal{C}_1$}


The next result ensures the satisfaction of the requirements of Theorem \ref{thm:stability-H} by formalizing the intuition that, when every solution to system $\mathcal{H}$ lie in $\mathcal{C}_1$ frequently enough, then desirable stability properties hold.  
\begin{thm}\label{thm:frequently-enough-in-C1} Given $T\in\Z_{\geq 1}$, consider system $\mathcal{H}$ and suppose there exist $\lambda_c:\mathcal{S}_2\to[0,1]$,  $\lambda_d:\mathcal{S}_2\to\Rlo$ with $\lambda_c\leq \lambda_d$,  $m_1,m_2:\mathcal{S}_2\to\Rlo$ and $\underline s,\overline s\in\Rlp$  such that the following holds for any solution $q$ and any $(k,j)\in\dom q$.
\begin{enumerate}[label=(\roman*)]
\item $\underline s I_{n_x}\preceq S(k,j) \preceq \overline s I_{n_x}$. 
\item $\sigma(a_1(k,j))\leq \lambda_c(q_2(0,0))$.
\item  
$\max\{\theta(q_2(k,j)),\nu_d(q_2(k,j))\}\leq \lambda_d(q_2(0,0))$ with $\nu_d$ and  $\theta$ as in Proposition \ref{prop:P1-P3}.
\item If $q(k,j)\notin \mathcal{C}_1$ and $\tau(k,j)=1$,  $L(\widehat X(k,j),X(k,j),U(k,j))\neq \emptyset$.
\item 
$(k-2j)\ln(\lambda_c(q_2(0,0)))+(3j+1)\ln(\lambda_d(q_2(0,0))) \leq m_1(q_2(0,0))-m_2(q_2(0,0))(k+j)$.
\end{enumerate}
Then $\mathcal{A}$ is stable. In addition, 
\begin{itemize}
\item if $m_1$ is constant, then $\mathcal{A}$ is uniformly stable,
\item if $m_2(q_2(0,0))<1$, then $\mathcal{A}$ is GAS,
\item if $m_1$, $m_2$ are constant and $m_2<1$, then $\mathcal{A}$ is UGES.\hfill $\Box$
\end{itemize}
\end{thm}

\noindent\textbf{Proof:} Let $q$ be a solution to $\mathcal{H}$ and $(k,j)\in\dom q$. By items (ii)-(iv) of Theorem \ref{thm:frequently-enough-in-C1} and (\ref{eq:pi-H}),   
$\pi(q,k,j)  \leq \lambda_c(q_2(0,0))^{k-2j}\lambda_d(q_2(0,0))^{3j+1}$. 
By item (v) of Theorem \ref{thm:frequently-enough-in-C1}, we derive
$\pi(q,k,j)  \leq  \dst \exp(m_1(q_2(0,0))-m_2(q_2(0,0))(k+j))$. 
By invoking item (i) of Theorem \ref{thm:frequently-enough-in-C1} and as $m_2(q_2(0,0))\geq 0$, we derive that item (i) of Theorem \ref{thm:stability-H} holds with $\mu(s,z)=\frac{\overline s}{\underline s}\exp(m_1(z))$ for any $(s,z)\in\Rlo\times\mathcal{S}_2$. Hence $\mathcal{A}$ is  stable by Theorem \ref{thm:stability-H}. The last three items of Theorem \ref{thm:frequently-enough-in-C1} follow by application of Theorem \ref{thm:stability-H}. \hfill $\blacksquare$

Item (i) of Theorem \ref{thm:frequently-enough-in-C1} states that $S$ is uniformly lower and upper-bounded by positive definite matrices for any solution $q$ to system $\mathcal{H}$, in which case it means that $\mathcal{S}_S\subset\{S\in\mathbb{S}^{n_x}_{\succ 0}\,:\, \underline s I_{n_x}\preceq S \preceq \overline s I_{n_x}\}$; a similar assumption is made in \cite{Liu_CDC2023}. Item (ii) upper-bounds the decay rate of $\mathcal{U}$ in view of Proposition \ref{prop:P1-P3} using $\lambda_c(q_2(0,0))$. This property is for instance verified  with $\lambda_c\equiv c$ when $\sigma(a_1)\leq c\in(0,1]$ along any solution to $\mathcal{H}$, which can be imposed by design. Item (iii), on the other hand, upper-bounds the growth rate of $\mathcal{U}$ using $\lambda_d$, which is typically  equal or larger than $1$. This property holds when $\theta$ admits a constant upper-bound, say $c'$, with $\lambda_d\equiv\max\{c',\overline s/\underline s\}$ as $\nu_d(q_2)\leq \overline s/\underline s$  by item (i) of Theorem \ref{thm:frequently-enough-in-C1}. Recall that we can relate $\theta$ to the parameters of Property \ref{property-lyap-ineq-initial-learning}, as shown in Lemma \ref{lem:theta}. Item (iv) means that, whenever the solution is not in $\mathcal{C}_1$ and an episode has not just occurred, it is possible to update the controller gain $K$; whereas this is the case in the numerical examples of Section \ref{sect:numerical-illustration}, this assumption may be restrictive and in general it cannot be guaranteed up-front due to the data collection happening in closed-loop and the time-varying nature of the problem. To mitigate this requirement, we plan to investigate the addition of exploratory signals to guarantee richness of the collected data. Finally, item (v) is inspired by \cite[Proposition 3.29]{Goebel-Sanfelice-Teel-book}, and is a condition that the decay rate of $\mathcal{U}$ on $\mathcal{C}_1$, which is captured by  $\lambda_c$,  compensates the growth of $\mathcal{U}$ when solutions leaves $\mathcal{C}_1$ described by  $\lambda_d$. This condition is thus satisfied when solutions remain sufficiently frequently in $\mathcal{C}_1$, and can be related to (average) dwell-time conditions that are customary in the switched systems literature. 


\section{Numerical illustration}\label{sect:numerical-illustration}

\subsection{Algorithmic implementation of $L$}\label{sect:numerical-implementation}
We discuss here the implementation details on the design of the map $L$ in (\ref{eq:L-example}). 
We propose solving an SDP with variables $\varsigma,Y,H$ subject to LMI constraints (\ref{eq:KDesign}) and with the objective of maximizing det($H$) to promote an increase of the matrices set to which the controller is robust by design. 
From the solution to this SDP we get 
the controller gain $K$ and the matrix $S$ defining the quadratic Lyapunov-like function $V(\cdot,S)$ as in Proposition \ref{prop:initial-learning-young}. To uniquely determine the remaining output of $L$, we fix $\epsilon_F \in (0,1)$ and obtain 
\begin{equation}\label{eq:KDesign-values}
\begin{aligned}
\hat{\varsigma}&=\tfrac{\varsigma}{\varsigma+1},\quad F=(1-\epsilon_F)\hat{\varsigma}H,\quad
a_1=1-a, \\
a_2 & =1+\varsigma^{-1},\quad
a=\max\left\{a'\in\Rlo\,:\, a' S^{-1} \preceq \epsilon_F H\right\}.
\end{aligned}
\end{equation}
The parameter $\epsilon_F$ influences the closed-loop properties we can conclude from this control design. Precisely, choosing smaller $\epsilon_F$ enlarges the set $\mathcal{E}(\data,F)$ at the cost of a lower decay rate, and viceversa. 
In all the analyses we selected $\epsilon_F=0.1$, and adaptively chose $\sigma(a_1)=1-0.1(1-a_1)$ consistently with Corollary \ref{cor:eventually-in-C1} and Theorem \ref{thm:frequently-enough-in-C1}. 
Unless otherwise stated, we study the problem of controlling LTV systems obtained from time-varying perturbations applied to the LTI system 
\begin{equation}\label{eq:plant-Ex1_LTI}
A_0 = 
 \begin{bmatrix}
  1.1 & 0.1 \\   
  0.1 & 0.2 \\   
 \end{bmatrix},\quad 
 B_0 =
 \begin{bmatrix}
  0.5 & 1 \\   
  0.1 & 0.2 \\   
 \end{bmatrix}, 
\end{equation}
which has one unstable mode but is stabilizable.
We consider a simulation horizon $[0,100]$ and, slightly departing from the theoretical framework\footnote{This could also have been done before time $(0,0)$, see Remark \ref{rem:initial-data-variables}.}, we impose a first episode of fixed length $[0,T], T=n_x+n_u$, 
where we excite the system with i.i.d uniformly distributed input with values in $[-1,1]$.
We denote hereafter by $K_0$ the control gain obtained at the end of this initial interval by evaluating map $L$. 
The SDPs were solved using MOSEK \cite{mosek} and the code is available\footnote{ 
https://github.com/col-tasas/2024-DD-adaptive-ETC-LTV}.

\subsection{Switching plant}\label{sect:numerical-illustration-Switching}
We consider an LTV plant that periodically switches between two stabilizable LTI systems with period $p$ 
\begin{equation}\label{eq:plant-Ex1}
\begin{array}{rlll}
A(k) \!\!\!\!& = A_0,\;\; \forall k \in \Zo,\vspace{0.1in}
 \\
B(k) \!\!\!\!& = \begin{cases}
 \begin{bmatrix}
  0.5 & 1 \\   
  0.1 & 0.2 \\   
 \end{bmatrix}, & \hspace{-0.2in} k\in [1+p(z-1),pz], z=2n+1, \vspace{0.1in} \\ 
\begin{bmatrix}
 0.5 & -\ell \\   
  0.1 & -0.2\ell \\   
 \end{bmatrix}, & \hspace{-0.1in} k\in [1+p(z-1),pz], z=2n, \\
\end{cases}
\end{array}
\end{equation}
where $n \in \Zo$, $p=12$ and $\ell =1$ unless otherwise stated. 
The switch 
is inspired by the case analyzed in Theorem \ref{thm:frequently-enough-in-C1} which certifies stability when the system remains sufficiently frequently in the region with Lyapunov function decrease. 
We compare in Figure \ref{fig:plant-Ex1} the proposed event-triggered scheme with the strategy of never updating the initial controller $K_0$ (fixed controller) and with a time-triggered strategy that triggers a new episode according to a preset period $n_p$. Specifically, we consider the oracle case where the time-triggered strategy has knowledge of the exact period (i.e., $n_p=p$) and
\emph{imperfect} scenarios where $n_p=p\pm4$. 
We also consider the fixed controller strategy for the case $\ell=2.5$. Square, circle and star markers denote plant switches, triggering instants and end of initial exploration phase, respectively. 
\begin{figure}[h]
    \centering
    \includegraphics[width=1\columnwidth]{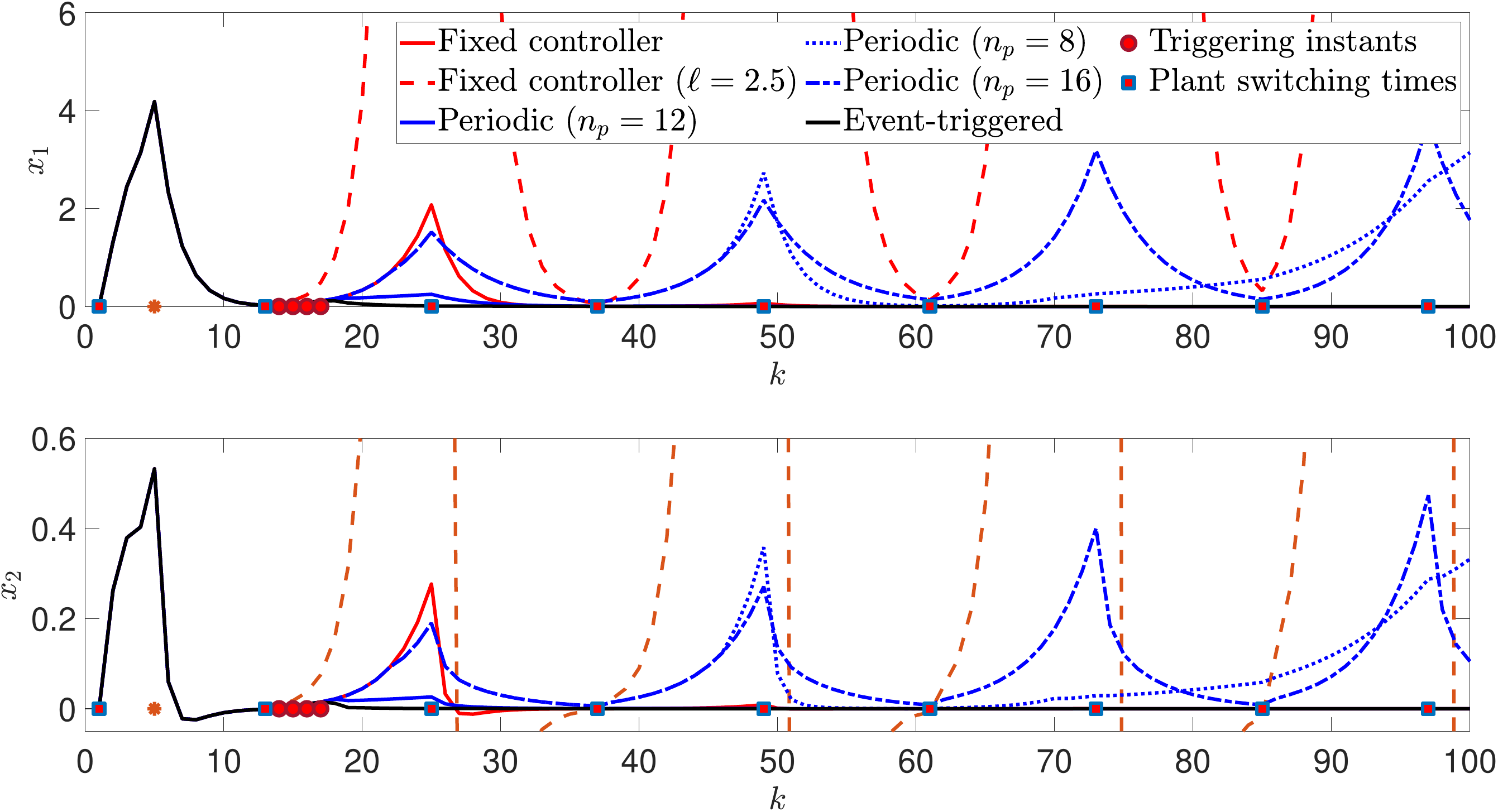}
    \caption{Closed-loop state response of (\ref{eq:plant-Ex1}) with different controllers (fixed, adaptive periodic time-triggered, and adaptive event-triggered).}
    \label{fig:plant-Ex1}
\end{figure}

The results show that the event-triggered scheme (solid black curve) outperforms all the other schemes at regulating the system around the origin. From the reported triggering instants it can be seen that the scheme only triggers a few times after the plant first switch has occurred and does this only using observed data. On the other hand, the time-triggered strategy shows low robustness to inexact choice of period. Indeed, while the performance of the oracle case (solid blue curve) is only slightly worse than the even-triggered case, the other two cases all display undesired oscillations and diverging behaviours. Note that the non-adaptive solution using $K_0$ (solid red curve) stabilizes the system when $\ell =1$, but leads to diverging response when the second column of the input matrix also changes magnitude ($\ell=2.5$).

\subsection{Sinusoidal variations}\label{sect:numerical-illustration-sinusoidal}
We consider now the following perturbations to (\ref{eq:plant-Ex1_LTI})
\begin{equation}\label{eq:plant-Ex2}
\begin{array}{rlll}
A(k) & = 
 A_0\left(I_2 +
 \delta_a\text{diag}( \cos{\frac{2\pi}{p}k},- \cos{\frac{2\pi}{p}k})
 \right),\vspace{0in} \\
B(k) &=
B_0 \; \forall k \in \Zo
\end{array}
\end{equation}
with $\delta_a=0.8$ unless otherwise stated. The input matrix is now constant and the state matrix undergoes a structured sinusoidal perturbation of period $p$ which determines a change up to $80 \%$ for the diagonal values. Figure \ref{fig:plant-Ex2} shows the closed-loop response of (\ref{eq:plant-Ex2}) 
with the proposed adaptive controller for different values of $p$.
\begin{figure}[h]
    \centering
    \includegraphics[width=1\columnwidth]{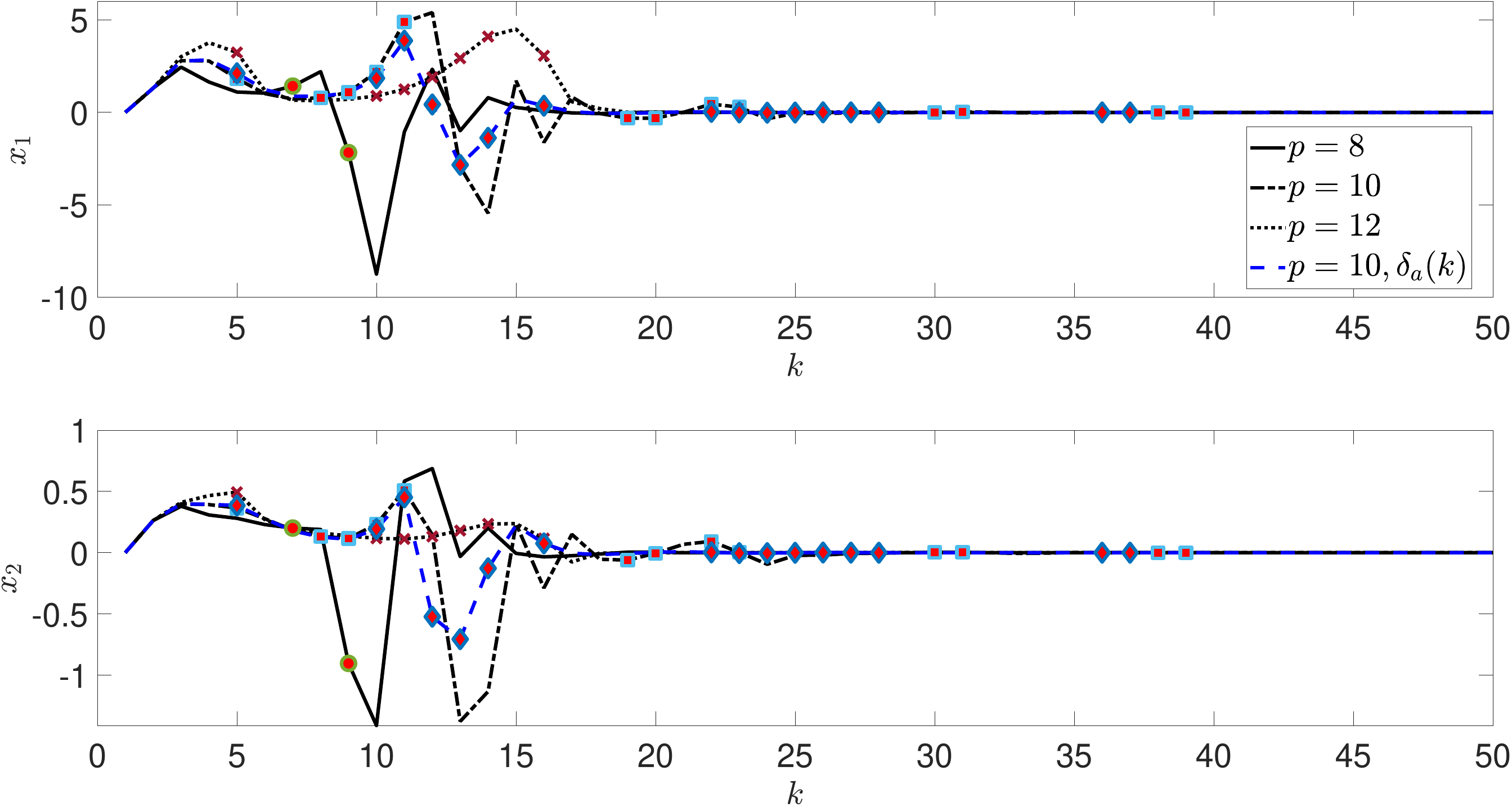} 
    \caption{Closed-loop state response of system (\ref{eq:plant-Ex2}) controlled with the adaptive event-triggered controller for different values of $p$ and $\delta_a$.  
    Triggering instants denoted by markers on the corresponding curve.
    }
    \label{fig:plant-Ex2}
\end{figure}

The results show that in all cases the event-triggered controllers successfully regulate the system to the origin with only a few episodes. The simulation was carried out until $k=100$ but the $x$-axis is stopped before as all curves reached the origin and no further episode is triggered. 
The last curve in Figure \ref{fig:plant-Ex2} refers to 
the case where $p=10$ and the parameter $\delta_a$ is time-varying and vanishes in finite-time $T_{\delta}=30$
\begin{equation}\label{eq:plant-Ex1-b}
\begin{array}{rlll}
\delta_a(k) & = \begin{cases}
 -\frac{1}{T_{\delta}}k+1, &  k\in [0, T_{\delta}], \vspace{0in} \\ 
0, & \hspace{-0.1in} k\geq T_{\delta}. \\
\end{cases}
\end{array}
\end{equation}
Under mild excitation conditions on the data matrices, we can invoke here Theorem \ref{thm:stability-eventually-in-C1} to give a-priori guarantees on the stability of the system. Take $k \geq T_{\delta}+T$. 
If the decrease condition (\ref{eq:episode-condition}) is always satisfied, then item (i) is automatically verified with $T^\star=T_{\delta}+T$.  
If not, but there exists a time step $\bar{T}$ where a new episode is triggered and map $L$ in (\ref{eq:L-example}) is non-empty\footnote{Because $(A_0,B_0)$ is stabilizable, a sufficient condition for non-emptiness of $L$ 
is that $Z=\left[\begin{smallmatrix}
        U \\ \widehat X
   \end{smallmatrix}\right]$ has full row rank.}, 
   then the resulting controller $K^+$ is guaranteed to stabilize $(A_0,B_0)$ because $\widehat X,X,U$ only contain data collected from an LTI system and thus $(A_0,B_0) \in \mathcal{E}(\data,F)$ for any $F \in \mathbb{S}^{n_x}_{\succ 0}$ (recall the discussion below (\ref{eq:D0-Delta})). 
Then item (i) is verified with $T^\star=\bar{T}$. Item (ii) holds by boundedness of the matrices (\ref{eq:plant-Ex2}) and of the controllers generated by the map $L$. 


\subsection{Comparison with time-triggered adaptation}\label{sect:numerical-illustration-comparison}
We finally compare our approach with the one 
in \cite{Liu_CDC2023} 
labeled ODDAC, which also considers the problem of data-based adaptive control of LTV systems. The system matrices are assumed to have a bounded rate of variation per timestep with known Lipschitz constant $L_c$. This knowledge is used to periodically design data-based state-feedback controllers which are robust to the predicted variations of the plant. We consider the same plant proposed in \cite{Liu_CDC2023}, which has $n_x=5$, $n_u=2$ and consists of a cubic interpolation of three LTI matrices with $L_c=0.0037$ inside an interval $[0,1000]$. We use the same algorithm parameters provided in \cite{Liu_CDC2023} except for choosing the period of control updates\footnote{This is the largest period for which we were able to find feasible solutions qualitatively matching those reported in \cite{Liu_CDC2023}.} $T_p=30$. In Figure \ref{fig:plant-Ex3} we compare it with our approach  by showing the Euclidean norm of the state. Note that at $k=0$ ODDAC is controlled with a precomputed stabilizing gain, whereas for the first $T=n_x+n_u$ timesteps our algorithm randomly explores as described in Section \ref{sect:numerical-implementation}. Besides the nominal scenario, we also consider the case where every entry of the state matrix $A(k)$ is uniformly scaled up by a factor $L_s\in\{1.1, 1.15, 1.2 \}$.
\begin{figure}[h]
    \centering
    \includegraphics[width=0.9\columnwidth]{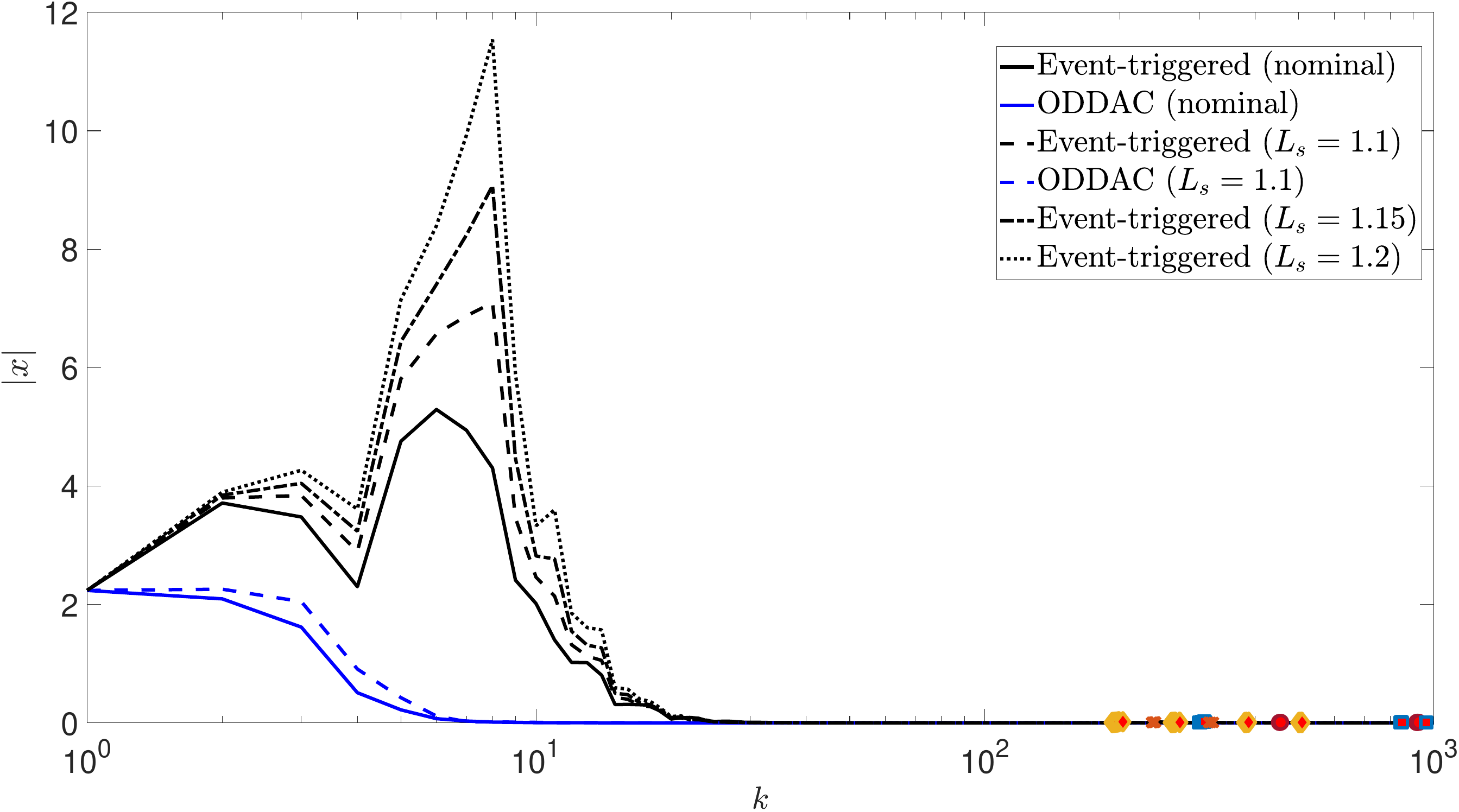} 
    \caption{Comparison between the proposed approach and ODDAC with plant from \cite{Liu_CDC2023}. Markers denote event-based controller updates.}
    \label{fig:plant-Ex3}
\end{figure}
\newline
The simulations show that ODDAC has good performance when the plant variations satisfy the assumed bounds $L_c$ and also possesses some degree of robustness to it (for $L_s=1.1$). However, it leads to unstable responses when the mismatch increases ($L_s\in\{1.15, 1.2 \}$) and the respective curves are not reported. Our method not only stabilizes the system with fewer controller updates (as shown by the markers in Figure \ref{fig:plant-Ex3}) and with no prior knowledge of the plant's variation, but notably does so also in the face of larger perturbations of the plant. 

\section{Conclusions}\label{sect:conclusions}

This work addresses the data-based stabilization of linear time-varying systems by on-line adaptation of the feedback gain. We propose a hybrid systems framework, which casts the adaptation as a jump in the dynamics triggered by events depending on some relevant closed-loop properties. The control design uses robust LMI conditions based on the most recently collected data to guarantee a Lyapunov-like property for all the systems sufficiently close to those that generated the data. 
The formulation allows the problem of establishing how and when adaptation should take place to be formally addressed, and a closed-loop analysis of the resulting nonlinear feedback loop to be performed.
We emphasize the prescriptive nature of the proposed framework, which accommodates various feedback design methods and triggering conditions under minor modifications.
It covers in a unified fashion various scenarios without relying on explicit conditions on the properties of the unknown plant matrices. A downside is that the connection between the presented stability conditions and the properties of the unknown model is not always obvious. Two case studies are provided to  shed more light on this key aspect, which we plan to further investigate in a future work by focusing on specific classes of LTV systems.

\section*{Acknowledgment}

We thank the anonymous reviewers for their insightful comments, which helped us strengthening the contribution.

\section*{Appendix: Proof of Proposition \ref{prop:initial-learning-young}}

We rely on the next intermediate lemma. 
\begin{lem}\label{lem:young} Under the conditions of Proposition \ref{prop:initial-learning-young}, for any $\varepsilon\in\Rlo$,
$\left[M_A \, \, M_B
\right]^\top\in \mathcal{E}(\data,F+\varepsilon S^{-1})$ implies 
$S - (M_A+M_B K)^{\top} \Big(\star\Big)^{-1}(M_A+M_B K)   \succ 0$, 
with $S=(\widehat X Y)^{-1}$, and $\star=S^{-1}-H+(1+\varsigma^{-1})(F+\varepsilon S^{-1})$. \hfill $\Box$
\end{lem}

\noindent\textbf{Proof sketch:}
We adapt the main steps of \cite[Theorem 5]{de-persis-tesi-tac2020} to the LTV context. We restrict the controller $K$ we search over to be of the form
\begin{equation}
\begin{array}{rllll}
\left[\begin{matrix}
        K \\ I_{n_x}
    \end{matrix}\right] & = & \left[\begin{matrix}
        U \\ \widehat{X}
    \end{matrix}\right]G,
\end{array}\label{eq:prop-relation-K-U0-X0}    
\end{equation}
with $G\in\R^{T\times n_x}$. For $M_A\in\R^{n_x\times n_x}, M_B\in\R^{n_x\times n_u}$, it holds 
$M_A + M_BK  =  (X + \widetilde D(q,M_A,M_B))G$. 
We then \change{consider that $M_A$ and $M_B$ satisfy  $\left[M_A \, \, M_B
\right]^\top\in \mathcal{E}(\data,F+\varepsilon S^{-1})$  and} show 
existence of $P\in \mathbb{S}^{n_x}_{\succ 0}$ and $\widetilde W \in\mathbb{S}^{n_x}$ such that
\begin{equation}
\begin{array}{rlllll}
(X + \widetilde D)G P G^\top (X + \widetilde D)^\top - P & \prec & -\widetilde W. 
\end{array}\label{eq:prop-alternative-lyap-condition}
\end{equation}
For this, we can upper bound the l.h.s. of (\ref{eq:prop-alternative-lyap-condition}) with the term
\begin{equation}\label{Lemma7_UB}
\varsigma XY(\widehat{X}  Y)^{-1} (X Y)^\top +  (1+\varsigma^{-1}) (F+\varepsilon P)  - \varsigma X X^\top - H.
\end{equation}
where we used: Young's inequality with $\varsigma \in \Rlp$; the fact that $\big[M_A \, \,M_B
\big]^\top\in \mathcal{E}(\data,F+\varepsilon P)$ for $\varepsilon\in\Rlo$, $F\in\mathbb{S}_{\succ 0}^{n_x}$; and the existence of $P :=\widehat{X} Y \in \mathbb{S}^{n_x}_{\succ 0}$ from the second LMI in (\ref{eq:KDesign}). In turn we can upper bound (\ref{Lemma7_UB}) with
$ -W+(1+\varsigma^{-1})\varepsilon P =: -\widetilde W$ by observing that $W:= H-(1+\varsigma^{-1}) F\succ 0 $ as $F\prec \frac{\varsigma}{1+\varsigma}H$, which yields (\ref{eq:prop-alternative-lyap-condition}). By Schur complementing (\ref{eq:prop-alternative-lyap-condition}) we obtain
\begin{equation}\label{eq:proof-young-main-intermediate-step}
P^{-1} - ((X + \widetilde D)G)^{\top} (P-\widetilde W)^{-1}(X + \widetilde D)G   \succ 0,
\end{equation}
where $P-\widetilde W \in \mathbb{S}^{n_x}_{\succ 0}$ from the first LMI in (\ref{eq:KDesign}). 
Taking $S=P^{-1}$, using $\big[M_A \, \,M_B
\big]^\top\in \mathcal{E}(\data,F+\varepsilon S^{-1})$ yield the result. 
\hfill $\blacksquare$

We are now ready to prove Proposition \ref{prop:initial-learning-young}.  Let $\varepsilon\in\Rlo$, $\left[M_A \, \, M_B
\right]^\top\in \mathcal{E}(\data,F+\varepsilon S^{-1})$ with $S$ as in Lemma \ref{lem:young}, and $x\in\R^{n_x}$. By Lemma \ref{lem:young},
$x^\top(M_A+M_B K)^{\top} (S^{-1}-\widetilde{W})^{-1} (M_A+M_B K)x - x^\top S x   \leq  0$ with $\widetilde W = H-(1+\varsigma^{-1})(F+ \varepsilon S^{-1})$ as in the proof of Lemma \ref{lem:young}, which we rewrite as
$x^\top(M_A+M_B K)^{\top} S (M_A+M_B K)x - x^\top S x   \leq 
- x^\top(M_A+M_B K)^{\top} \big[\big(S^{-1}-\widetilde W\big)^{-1}-S\big](M_A+M_B K)x$. 
By taking $V(x,S)=x^{\top}Sx$ and $Q:=\big(S^{-1}-\widetilde W\big)^{-1}-S$, 
\begin{equation}\label{eq:young-second-main-intermediate-step}
\begin{aligned}
&V((M_A+M_B K)x,S) - V(x,S)  \leq \\
&\quad - x^\top(M_A+M_B K)^{\top}Q(M_A+M_B K)x. 
\end{aligned}
\end{equation}

We show next that there exists $\lambda(\varepsilon) \in (-1,+\infty)$ such that $Q \succeq \lambda(\varepsilon) S$. We  note that $a=\max\big\{a'\in\Rlo\,:\, a' S^{-1} \preceq  W\big\}$ belongs to $(0,1)$ as $S, W \in \mathbb{S}^{n_x}_{\succ 0}$ and $P \succ H$ (from (\ref{eq:KDesign})) implies $S^{-1}- W \succ 0$. Take
\begin{equation}\label{eq:lambda}
\begin{array}{rlll}
\lambda(\varepsilon):=\frac{a-(1+\varsigma^{-1}) \varepsilon}{1-a+(1+\varsigma^{-1}) \varepsilon},
\end{array}
\end{equation}
where the denominator is positive because $\varepsilon \in \Rlo$,  $a \in (0,1)$ and $\varsigma>0$.  Moreover, it holds
\begin{equation}\label{eq:lambda-property}
\begin{array}{rll}\lambda(\varepsilon)+1 
=\frac{1}{1-a+(1+\varsigma^{-1}) \varepsilon}>0.
\end{array}
\end{equation}
Hence $\lambda(\varepsilon) \in (-1,+\infty)$. To show $Q \succeq \lambda(\varepsilon) S$, observe that
\begin{equation}
\begin{aligned}
Q \succeq \lambda(\varepsilon) S  &\Leftrightarrow  \left(S^{-1}-W+(1+\varsigma^{-1}) \varepsilon S^{-1}\right)^{-1}\!-\!S\succeq \lambda(\varepsilon) S \\
&\Leftrightarrow S^{-1}\!-\!W\!+\! (1+\varsigma^{-1})\varepsilon S^{-1} \!\!\preceq \tfrac{1}{1+ \lambda(\varepsilon)} S^{-1} \\
&\Leftrightarrow \big(1-\tfrac{1}{1+\lambda(\varepsilon)}+(1+\varsigma^{-1}) \varepsilon \big) S^{-1} \preceq W \\
&\Leftrightarrow \big(\tfrac{\lambda(\varepsilon)}{1+\lambda(\varepsilon)}+(1+\varsigma^{-1}) \varepsilon \big) S^{-1} \preceq W. \\
\end{aligned}
\end{equation}
By (\ref{eq:lambda}) and (\ref{eq:lambda-property}), $\tfrac{\lambda(\varepsilon)}{1+\lambda(\varepsilon)}+(1+\varsigma^{-1}) \varepsilon=a$ therefore
$Q \succeq \lambda(\varepsilon) S  \Leftrightarrow a S^{-1} \preceq W$. 
The last inequality holds by definition of $a$. 
We deduce that $Q \succeq \lambda(\varepsilon) S$. We use this property together with (\ref{eq:young-second-main-intermediate-step}) to derive that for all $\left[M_A \, \,M_B
\right]^\top\in \mathcal{E}(\data,F+\varepsilon S^{-1})$, 
$V((M_A+M_B K)x,S) - V(x,S) \leq - \lambda(\varepsilon) V((M_A+M_B K)x,S)$ 
and thus, by (\ref{eq:lambda-property})
$V((M_A+M_B K)x,S)  \leq \frac{1}{1+\lambda(\varepsilon)}  V(x,S) =(1-a+(1+\varsigma^{-1}) \varepsilon)V(x,S)$. The desired result holds with 
 $a_1=1-a\in(0,1)$ and $a_2=1+\varsigma^{-1}$.

\bibliographystyle{ieeetr}
\bibliography{bib_global.bib}

\begin{thebibliography}{10}

\bibitem{sanfelice-teel-aut10}
R.~Sanfelice and A.~Teel, ``Dynamical properties of hybrid systems
  simulators,'' {\em Automatica}, vol.~46, no.~2, pp.~239--248, 2010.

\bibitem{Heemels-Johansson-Tabuada-cdc12}
W.~Heemels, K.~Johansson, and P.~Tabuada, ``An introduction to event-triggered
  and self-triggered control,'' in {\em IEEE Conference on Decision and
  Control}, pp.~3270--3285, 2012.

\bibitem{de-persis-tesi-tac2020}
C.~{De Persis} and P.~Tesi, ``Formulas for data-driven control: Stabilization,
  optimality, and robustness,'' {\em IEEE Trans. on Automatic Control},
  vol.~65, no.~3, pp.~909--924, 2020.

\bibitem{van-waarde-et-al-tac2020(data)}
H.~{van Waarde}, J.~Eising, H.~Trentelman, and M.~Camlibel, ``Data
  informativity: a new perspective on data-driven analysis and control,'' {\em
  IEEE Trans. on Aut. Control}, vol.~65, no.~11, pp.~4753--4768, 2020.

\bibitem{rotulo-et-al-aut22}
M.~Rotulo, C.~{De Persis}, and P.~Tesi, ``Online learning of data-driven
  controllers for unknown switched linear systems,'' {\em Automatica},
  vol.~145, p.~110519, 2022.

\bibitem{Eising_CDC22_DDswitched}
J.~Eising, S.~Liu, S.~Mart\'inez, and J.~Cort\'es, ``Using data informativity
  for online stabilization of unknown switched linear systems,'' in {\em IEEE
  Conference on Decision and Control}, 2022.

\bibitem{Liu_CDC2023}
S.~Liu, K.~Chen, and J.~Eising, ``Online data-driven adaptive control for
  unknown linear time-varying systems,'' in {\em IEEE Conference on Decision
  and Control}, 2023.

\bibitem{nortmann-mylvaganam-tac23}
B.~Nortmann and T.~Mylvaganam, ``Direct data-driven control of linear
  time-varying systems,'' {\em IEEE Trans. on Automatic Control}, vol.~68,
  no.~8, pp.~4888--4895, 2023.

\bibitem{Mazo-Anta-Tabuada-Aut10}
M.~Mazo, A.~Anta, and P.~Tabuada, ``{An ISS self-triggered implementation of
  linear controllers},'' {\em Automatica}, vol.~46, no.~8, pp.~1310--1314,
  2010.

\bibitem{Wang-Lemmon-aut11}
X.~Wang and M.~Lemmon, ``On event design in event-triggered feedback systems,''
  {\em Automatica}, vol.~47, no.~10, pp.~2319--2322, 2011.

\bibitem{maass-et-al-tac2023(etc)}
A.~Maass, W.~Wang, D.~Ne{\v{s}}i{\'c}, R.~Postoyan, and W.~Heemels,
  ``Event-triggered control through the eyes of a hybrid small-gain theorem,''
  {\em IEEE Trans. on Automatic Control}, vol.~68, no.~10, pp.~5906--5921,
  2023.

\bibitem{de-persis-et-al-tac2023(etc-data)}
C.~{De Persis}, R.~Postoyan, and P.~Tesi, ``Event-triggered control from
  data,'' {\em IEEE Trans. on Automatic Control}, vol.~available on-line, 2023.

\bibitem{qi2022data}
W.-L. Qi, K.-Z. Liu, R.~Wang, and X.-M. Sun, ``Data-driven
  $\mathcal{L}_{2}$-stability analysis for dynamic event-triggered networked
  control systems: A hybrid system approach,'' {\em IEEE Trans. on Industrial
  Electronics}, vol.~70, no.~6, pp.~6151--6158, 2022.

\bibitem{wang-et-al-tc23}
X.~Wang, J.~Berberich, J.~Sun, G.~Wang, F.~Allg{\"o}wer, and J.~Chen,
  ``Model-based and data-driven control of event-and self-triggered
  discrete-time linear systems,'' {\em IEEE Trans. on Cyb.}, vol.~53, no.~9,
  pp.~6066 -- 6079, 2023.

\bibitem{Goebel-Sanfelice-Teel-book}
R.~Goebel, R.~Sanfelice, and A.~Teel, {\em Hybrid Dynamical Systems}.
\newblock Princeton University Press, Princeton, U.S.A., 2012.

\bibitem{Girard-tac15}
A.~Girard, ``Dynamic triggering mechanisms for event-triggered control,'' {\em
  IEEE Trans. on Automatic Control}, vol.~60, no.~7, pp.~1992--1997, 2015.

\bibitem{sif-et-al-aut2023}
S.~Benahmed, R.~Postoyan, M.~Granzotto, L.~Bu{\c{s}}oniu, J.~Daafouz, and
  D.~Ne{\v{s}}i{\'c}, ``Stability analysis of optimal control problems with
  time-dependent costs,'' {\em Automatica}, vol.~157, p.~111272, 2023.

\bibitem{bisoffi-et-al-aut22}
A.~Bisoffi, C.~{De Persis}, and P.~Tesi, ``{Data-driven control via Petersen's
  lemma},'' {\em Automatica}, vol.~145, p.~110537, 2022.

\bibitem{vanWaarde_SICO23_QMI}
H.~{van Waarde}, K.~Camlibel, J.~Eising, and H.~Trentelman, ``Quadratic matrix
  inequalities with applications to data-based control,'' {\em SIAM J. on
  Contr. and Optim.}, vol.~61, no.~4, pp.~2251--2281, 2023.

\bibitem{borgers-2018(time-reg-etc)}
D.~Borgers, V.~Dolk, G.~Dullerud, A.~Teel, and W.~Heemels, ``Time-regularized
  and periodic event-triggered control for linear systems,'' in {\em Control
  Subject to Computational and Communication Constraints}, pp.~121--149,
  Springer, 2018.

\bibitem{Abdelrahim-et-al-aut17}
M.~Abdelrahim, R.~Postoyan, J.~Daafouz, and D.~Ne\v{s}i\'c, ``{Robust
  event-triggered output feedback controllers for nonlinear systems},'' {\em
  Automatica}, vol.~75, pp.~96--108, 2017.

\bibitem{tallapragada-chopra-cdc2012}
P.~Tallapragada and N.~Chopra, ``{Event-triggered dynamic output feedback
  control for LTI systems},'' in {\em IEEE Conference on Decision and Control},
  2012.

\bibitem{wang-et-al-cdc2019}
W.~Wang, D.~Ne{\v{s}}i{\'c}, R.~Postoyan, I.~Shames, and W.~Heemels,
  ``State-feedback event-holding control for nonlinear systems,'' in {\em IEEE
  Conference on Decision and Control}, 2019.

\bibitem{mosek}
M.~ApS, {\em The MOSEK optimization toolbox for MATLAB manual. Version 9.0.},
  2019.

\end{thebibliography}

\begin{IEEEbiography}[{\includegraphics[width=1in,height=1.25in,clip,keepaspectratio]{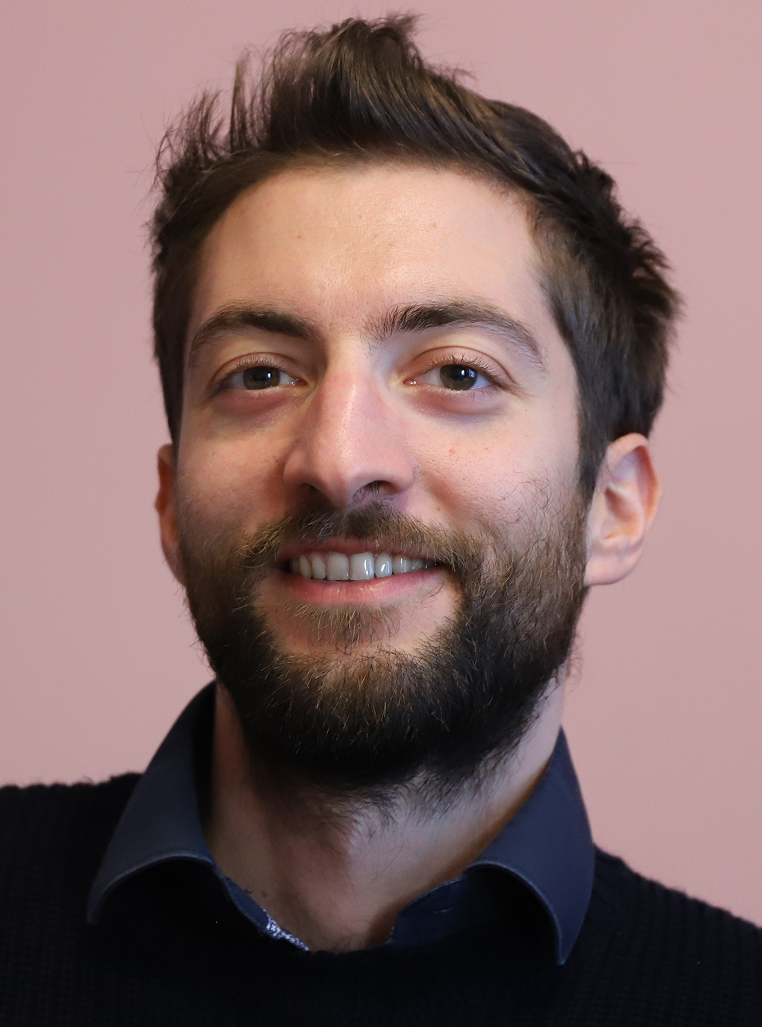}}]
{Andrea Iannelli} (Member, IEEE) is an Assistant Professor in the Institute for Systems Theory and Automatic Control at the University of Stuttgart. He completed his B.Sc. and M.Sc. degrees in Aerospace Engineering at the University of Pisa and received his PhD from the University of Bristol. He was also a postdoctoral researcher in the Automatic Control Laboratory at ETH Zurich. His main research interests are centered around
robust and adaptive control, uncertainty quantification, and sequential decision-making. 
    \end{IEEEbiography}
\vspace{-0.2in}
\begin{IEEEbiography}[{\includegraphics[width=1in,height=1.25in,clip,keepaspectratio]{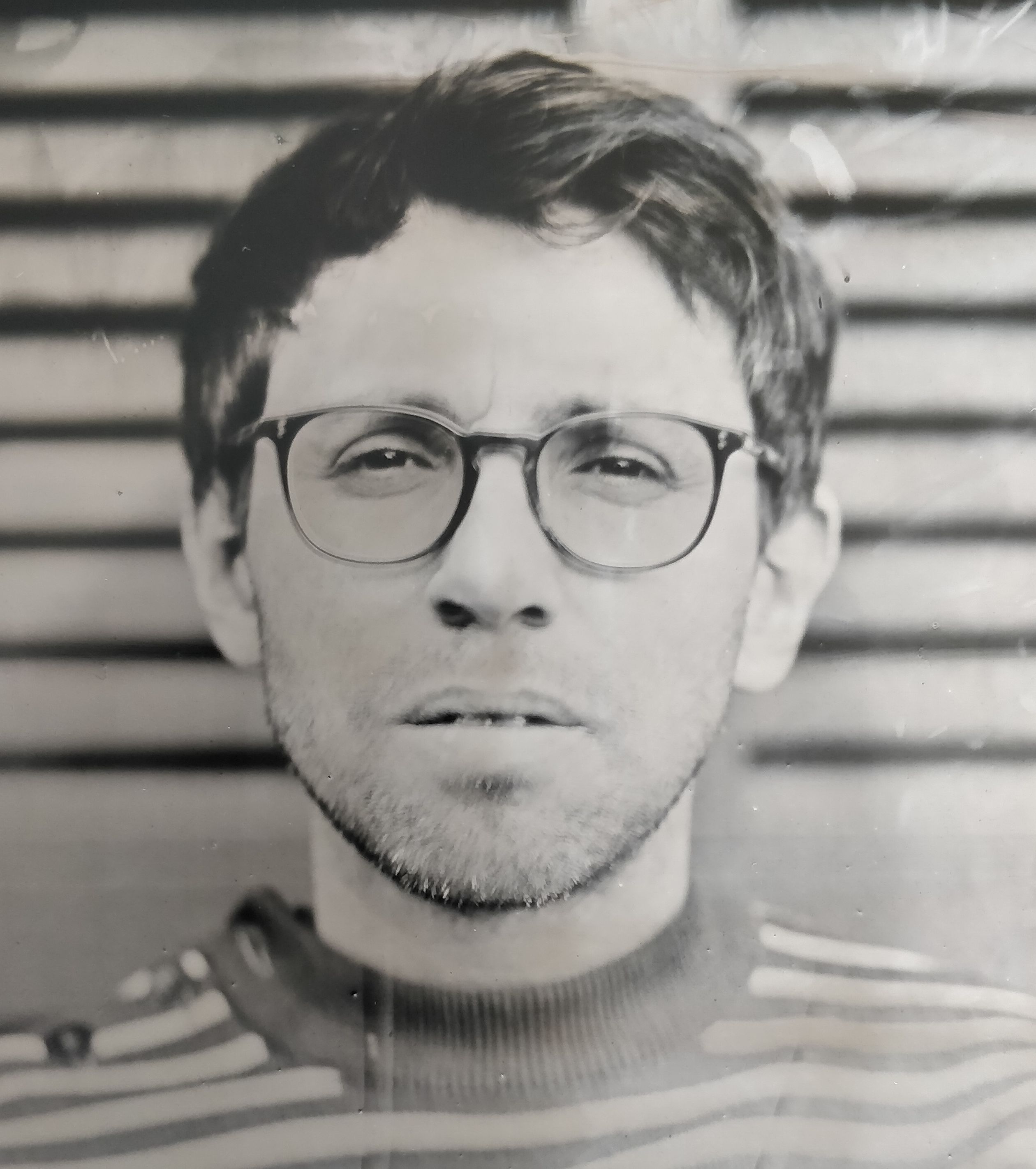}}]
{Romain Postoyan} (Senior Member, IEEE) got the ``Ing\'enieur'' degree in Electrical and Control Engineering from ENSEEIHT (France) in 2005, the M.Sc. by Research in Control Theory \& Application from Coventry University (United Kingdom) in 2006, and the Ph.D. in Control Engineering from Universit\'e Paris-Sud (France) in 2009. In 2010, he was a post-doc at the University of Melbourne (Australia). Since 2011, he is a CNRS researcher at CRAN (France). 
\end{IEEEbiography}

\end{document}